\documentclass[aps,prd,showpacs,twocolumn,groupedaddress,floatfix,amsfonts]{revtex4}
\usepackage{graphicx}
\usepackage{dcolumn}
\usepackage{bm}
\usepackage{amssymb}
\usepackage{amsmath}
\usepackage{xspace}
\usepackage{color}
\newcommand{\Ptg}{p_{T}^{\gamma}}

\newcommand{\ve}{\varepsilon}

\newcommand{\la}{\langle}
\newcommand{\ra}{\rangle}
\newcommand{\lt}{\!<\!}
\newcommand{\gt}{\!>\!}

\newcommand{\gpj}{$\gamma+{\rm jets}$\xspace}
\newcommand{\gpTHRj}{$\gamma+{\rm 3~jets}$\xspace}

\newcommand{\ptsj}{p_T^{\rm jet2}}
\newcommand{\dSphi}{$\Delta S_{\phi}$~}
\newcommand{\dSpt}{$\Delta S_{p_{T}}$~}
\newcommand{\dSptp}{$\Delta S_{p_{T}^\prime}$~}

\newcommand{\epsdi}{\varepsilon_{\rm DI}}
\newcommand{\epsdp}{\varepsilon_{\rm DP}}

\newcommand{\TeV}{\ensuremath{\text{TeV}}\xspace}

\begin{document}
\hspace{5.2in} \mbox{FERMILAB-PUB-09-644-E}
\title{\boldmath Double parton interactions in $\gamma$+3 jet events in $p\bar{p}$ collisions 
at $\sqrt{s}=1.96$~TeV}
% LIST_OF_AUTHORS_R2.TEX                 12/15/09           
%
\author{V.M.~Abazov$^{37}$}
\author{B.~Abbott$^{75}$}
\author{M.~Abolins$^{64}$}
\author{B.S.~Acharya$^{30}$}
\author{M.~Adams$^{50}$}
\author{T.~Adams$^{48}$}
\author{E.~Aguilo$^{6}$}
\author{G.D.~Alexeev$^{37}$}
\author{G.~Alkhazov$^{41}$}
\author{A.~Alton$^{64,a}$}
% ...

\affiliation{\vspace{0.1 in}(The D\O\ Collaboration)\vspace{0.1 in}}
% ...
\affiliation{$^{6}$Simon Fraser University, Burnaby, British Columbia, Canada;
                and York University, Toronto, Ontario, Canada}
% ...
\affiliation{$^{30}$Tata Institute of Fundamental Research, Mumbai, India}
% ...
\affiliation{$^{37}$Joint Institute for Nuclear Research, Dubna, Russia}
% ...
\affiliation{$^{41}$Petersburg Nuclear Physics Institute,
                St. Petersburg, Russia}
% ...
\affiliation{$^{48}$Florida State University, Tallahassee, Florida 32306, USA}
% ...
\affiliation{$^{50}$University of Illinois at Chicago,
                Chicago, Illinois 60607, USA}
% ...
\affiliation{$^{64}$Michigan State University,
                East Lansing, Michigan 48824, USA}
% ...
\affiliation{$^{75}$University of Oklahoma, Norman, Oklahoma 73019, USA}
% ...
  
\date{\today}

\begin{abstract}
     We have used a sample of $\gamma + 3$ jets events collected by the D0 experiment 
     with an integrated luminosity of about 1~fb$^{-1}$ 
     to determine the fraction of events with double parton %(DP) 
     scattering
     ($f_{\rm DP}$) in a single $p \bar{p}$ collision at $\sqrt{s}=1.96$ TeV.
      The DP fraction and  effective cross section ($\sigma_{\rm eff}$),
      a process-independent scale parameter related to the parton density
      inside the nucleon, are measured in three intervals
      of the second (ordered in $p_T$) jet transverse momentum $\ptsj$
      within the range $15 \leq \ptsj \leq 30$~GeV. 
      In this range, $f_{\rm DP}$ varies between $0.23 \leq f_{\rm DP} \leq 0.47$,
      while $\sigma_{\rm eff}$ has the average value
      $\sigma_{\rm eff}^{\rm ave} = 16.4 \pm 0.3(\rm stat) \pm 2.3(\rm syst)$ mb.

\end{abstract}
\maketitle

%%%%%%%%%%%%%%%%%%%%%%%%%%%%%%%%%%%%%%%%%%%%%%%%%%%%%%%%%%%%%%%%%%%%%%%%%%%%%%%%%%%%%%%
%%% Main text starts here

%\setpagewiselinenumbers
%\linenumbers

\section{Introduction}
\label{Sec:Intro}

\begin{figure}[h]
~\\[-2mm]
\hspace*{-2mm} \includegraphics[scale=0.30]{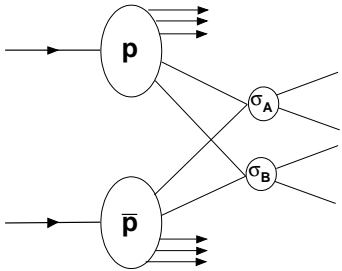}
~\\[-7mm]
\caption{Diagram of a double parton scattering event.}
\label{fig:dp_proc}
\vskip -5mm
\end{figure}

   Many features of high energy inelastic hadron collisions 
  depend directly on the parton structure of hadrons.
  The inelastic scattering of nucleons 
  need not to occur only through a single parton-parton
  interaction and the contribution from double 
  parton (DP) collisions 
  can be significant.
  A schematic view of a double parton scattering event in a $p\bar{p}$ interaction is shown
  in Fig.~\ref{fig:dp_proc}.
  The rate of events with multiple parton scatterings 
   depends on how the partons are spatially 
   distributed within the nucleon. 
Theoretical discussions and estimations~\cite{Landsh,Goebel,TH1,TH2,TH3} 
  stimulated measurements~\cite{AFS,UA2,CDF93,CDF97} of DP event fractions
 and DP cross sections. 
 The latter  can be expressed as  
  \begin{eqnarray}
   \sigma_{\rm DP} \equiv
     m \frac{\sigma_{\rm A} \sigma_{\rm B}}{2\sigma_{\rm eff} },
   \label{eq:sigma_DPS}
   \end{eqnarray} 
   where $\sigma_{\rm A}$ and $\sigma_{\rm B}$ are the cross sections
   of two independent partonic scatterings $A$ and $B$.
   The factor $m$   is equal to unity 
   when processes $A$ and $B$ are indistinguishable while  $m=2$  otherwise~\cite{TH3,Sjost,Threl}. 
  The process-independent scaling parameter
 $\sigma_{\rm eff}$ has the units of cross section. Its relation to the 
  spatial distribution of partons within the proton has been discussed 
  in~\cite{Landsh,TH1,TH2,Sjost,TH3,Threl}.
   The ratio $\sigma_B/\sigma_{\rm eff}$ can be interpreted as the probability
   for partonic process $B$ to occur provided that process $A$ has already occurred.
   If the partons are uniformly distributed inside the 
   nucleon (large $\sigma_{\rm eff}$), $\sigma_{\rm DP}$ will 
   be rather small and, conversely,  it will be large for a highly 
   concentrated parton spatial density (small $\sigma_{\rm eff}$).
   The implication and possible correlations of parton momenta distribution functions in (\ref{eq:sigma_DPS})
   are discussed in~\cite{Snigir,Stirling09,Sjost_JI}.

   In addition to constraining predictions from various models of nucleon
   structure and providing a better understanding of
   non-perturbative QCD dynamics, measurements of $f_{\rm DP}$
   and $\sigma_{\rm eff}$ are also needed for the accurate estimation of
   backgrounds for many rare new physics processes as well as for Higgs boson searches at
   the Tevatron and LHC~\cite{WH,Huss}.

   To date, there have been only four dedicated  measurements studying double parton scattering:
   by the AFS Collaboration in $pp$ collisions at $\sqrt{s}=63$~GeV~\cite{AFS},
   by the UA2 Collaboration in $p\bar{p}$ collisions at $\sqrt{s}=630$~GeV~\cite{UA2},
   and twice by the CDF Collaboration in $p\bar{p}$ collisions at $\sqrt{s}=1.8$ TeV~\cite{CDF93,CDF97}.
   The four-jet final state was used in the measurements to extract values of 
   $\sigma_{\rm DP}$ and then $\sigma_{\rm eff}$, and the \gpTHRj final state 
   was used in~\cite{CDF97} to extract $f_{\rm DP}$ fractions and then $\sigma_{\rm eff}$.
   The obtained values of $\sigma_{\rm eff}$ by those experiments are $\sigma_{\rm eff} \approx 5$~mb (AFS), 
   $\sigma_{\rm eff} \gt 8.3$~mb at the $95\%$ C.L. (UA2), $\sigma_{\rm eff}= 12.1^{+10.7}_{-5.4}$~mb (CDF, four-jet)
   and $\sigma_{\rm eff}= 14.5 \pm 1.7^{+1.7}_{-2.3}$~mb (CDF, \gpTHRj).
   Table~\ref{tab:sigeff_world} summarizes all previous measurements of $\sigma_{\rm eff}$, $\sigma_{\rm DP}$, 
   and $f_{\rm DP}$.

   This paper presents an analysis of hard inelastic
   events with a photon candidate (denoted below as $\gamma$) and at least 3 jets  %$\gamma+$3 jets+X 
   (referred to below as ``\gpTHRj'' events) collected 
   with the D0 detector~\cite{D0_det} at the Fermilab Tevatron Collider with
   $\sqrt{s}=1.96~\TeV$ and an integrated luminosity of 1.02 $\pm$ 0.06 fb$^{-1}$.
   In this final state, DP events are caused by two partonic scatterings,
   with $\gamma+$jets production in the first scattering and dijet production in the second.
   Thus, the rate of \gpTHRj events and their kinematics should be
   sensitive to a contribution from additional parton interactions.
   Differences in the types of the two final states ($\gamma+$jets and dijets)
   and better energy measurement of photons as compared
   with jets facilitate differentiation between the two DP scatterings as compared
   with the 4-jet measurements.
   Also,  it was shown in~\cite{Tao} that a larger fraction of DP events is expected
   in the \gpTHRj final state as compared with the 4-jet events.
   The large integrated luminosity 
   allows us to select \gpTHRj events at
   high photon transverse momentum,
   $60<\Ptg<80$~GeV (vs. $\Ptg>16$~GeV in CDF~\cite{CDF97}), with a larger photon
   purity~\cite{gamjet_PLB}. 
   The choice of a high threshold on the photon momentum provides 
   (a) a clean separation between the jet produced in the same parton scattering
   from which the photon originates and the jets originating from additional
   parton scatterings and 
   (b) a better determination of the energy scale of the \gpj process.
   Also, in contrast to~\cite{CDF97},
   the jet transverse momenta are corrected to the particle level. 
   Other differences in the technique used for extracting $\sigma_{\rm eff}$ are described below.
\begin{table*}[]
\vskip2mm
\centering
\small
\caption{Summary of the results, experimental parameters, and event selections
for the double parton analyses performed by the AFS, UA2 and CDF Collaborations.}
\label{tab:sigeff_world}
\vskip 1mm
\begin{tabular}{cccllll} \hline\hline
Experiment & $ \sqrt{s}$ (GeV) & Final state & $p_{T}^{min}$ (GeV) & $\eta $ range & ~~~~~$\sigma_{\rm eff}$ & ~~~~~~$\sigma_{\rm DP}$, $f_{\rm DP}$\\\hline
    AFS ($pp$), 1986~\cite{AFS}  & $ 63 $    & $ 4~{\rm jets} $ &  $p_{\rm T}^{\rm jet}>4$  & $|\eta^{\rm jet}|<1$  & $\sim 5$ mb & $\sigma_{\rm DP}/\sigma_{\rm dijet} = (6\pm1.5\pm2.0)\%$ \\\hline
    UA2 ($p\bar{p}$), 1991~\cite{UA2}   & $ 630 $   & $ 4~{\rm jets} $ &  $p_{\rm T}^{\rm jet}>15$ & $|\eta^{\rm jet}|<2$  & $> 8.3$ mb (95\% C.L.) & $ \sigma_{\rm DP}=0.49\pm0.20$ nb \\\hline
    CDF ($p\bar{p}$), 1993~\cite{CDF93}   & $ 1800 $  & $ 4~{\rm jets} $ &  $p_{\rm T}^{\rm jet}>25$ & $|\eta^{\rm jet}|<3.5$& $12.1_{-5.4}^{+10.7}$ mb & $\sigma_{\rm DP} = (63^{+32}_{-28})$ nb, $f_{\rm DP} = (5.4_{-2.0}^{+1.6})$\% \\\hline
    CDF ($p\bar{p}$), 1997~\cite{CDF97}   & $ 1800 $  & $ \gamma + 3~{\rm jets} $   &  $p_{\rm T}^{\rm jet}>6$ & $|\eta^{\rm jet}|<3.5$ & $   $ & \\
               &           &                  &  $p_{\rm T}^{\gamma}>16$  & $|\eta^{\gamma}|<0.9$ & $14.5$$\pm$$1.7_{-2.3}^{+1.7}$ mb & $ f_{\rm DP} = (52.6\pm 2.5\pm 0.9)$\% \\\hline\hline
\end{tabular}
\vskip 5mm
\end{table*}

   This paper is organized as follows. Section \ref{Sec:Method} briefly describes the technique
   used to extract the $\sigma_{\rm eff}$ parameter. Section \ref{Sec:ObjectID} provides the 
   description of the data samples and selection criteria. Section \ref{Sec:Models} describes
   the models used for signal and background events. 
   In Section \ref{Sec:Vars} we introduce the variables which allow us to distinguish DP events 
   from other $\gamma+3$ jets events and determine their fraction.
   The procedure for finding the fractions of DP events is described in Section \ref{Sec:Frac}.
   Section \ref{Sec:Eff} describes the determination of other parameters needed
   to calculate $\sigma_{\rm eff}$.
   Results of the measurement are given in Section \ref{Sec:Sigma_eff}
   with their application to selected models of parton density.

\section{Technique for extracting $\sigma_{\rm eff}$ from data}
\label{Sec:Method}

    In the 4-jet analyses~\cite{AFS,UA2,CDF93}, $\sigma_{\rm eff}$ was extracted
    from measured DP cross sections using Monte Carlo (MC) modeling for signal and
    background events and QCD predictions for the dijet cross sections.    
    Both MC modeling and the QCD predictions suffer from substantial uncertainties
    leading to analogous uncertainties in $\sigma_{\rm eff}$.
    Another technique for extracting  $\sigma_{\rm eff}$ was proposed in~\cite{CDF97}. 
    It uses only quantities
    determined from data and thus minimizes the impact of theoretical assumptions. 
    Here we follow this method 
    and extract $\sigma_{\rm eff}$ without theoretical predictions of the \gpj and dijets cross sections
    by comparing the number of  \gpTHRj events produced in DP interactions
    in single $p\bar{p}$ collisions to the number of \gpTHRj events produced in two distinct
    hard interactions occurring in two separate $p\bar{p}$ collisions in the same beam crossing.
    The latter class of events is referred to as double interaction (DI) events.  
    Assuming uncorrelated parton scatterings in the DP process~\cite{Landsh,Goebel,TH1,TH2,TH3,Threl},
    DP and DI events should be kinematically identical. 
    This assumption is discussed in Appendix A.

    Measurements of dijet production with jet $p_T\gtrsim 12-15$~GeV~\cite{D0_diff}
    in both central and forward rapidity~\cite{rapidity} regions indicate that 
    the contribution from single and double diffraction events represents $\lesssim 1\%$
    of the total dijet cross section. Therefore \gpj and dijet events with jet $p_T>15$~GeV
    are produced predominantly as a result of inelastic non-diffractive (hard) $p\bar{p}$ 
    interactions. In a $p\bar{p}$ beam crossing with two hard collisions
    the probability for a DI event in that crossing can be expressed as
    \begin{equation}
    P_{\rm DI} = 
    ~2 \frac{\sigma^{\gamma j}}{\sigma_{\rm hard}} \frac{\sigma^{jj}}{\sigma_{\rm hard}}.
    \end{equation} 
    Here ${\sigma^{\gamma j}}$ and ${\sigma^{jj}}$ are the cross sections to produce 
    the inclusive \gpj 
    and dijet events, which combined give the \gpTHRj final state, and $\sigma_{\rm hard}$ is the total
    hard $p\bar{p}$ interaction cross section.
    The factor $2$ takes into account that the two hard scatterings, producing a \gpj or dijet event, 
    can be ordered in two ways with respect to the two collision vertices in the DI events.
    The number of DI events, $N_{\rm DI}$, 
    can be obtained
    from $P_{\rm DI}$, after correction for the efficiencies to pass
    geometric and kinematic selection criteria $\epsilon_{\rm DI}$,
    the two-vertex event selection efficiency, $\epsilon_{\rm 2vtx}$,
    and the number of beam crossings with two hard collisions, $N_{\rm 2coll}$:
    \begin{eqnarray}    
    N_{\rm DI} = 
    ~2 \frac{\sigma^{\gamma j}}{\sigma_{\rm hard}} \frac{\sigma^{jj}}{\sigma_{\rm hard}}
    ~N_{\rm 2coll} ~\epsilon_{\rm DI} ~\epsilon_{\rm 2vtx}.
    \label{eq:ndi}
    \end{eqnarray}     

    Analogously to $P_{\rm DI}$, the probability for DP events, $P_{\rm DP}$, 
    in a beam crossing with one hard collision, is
    \begin{equation}
    P_{\rm DP} = \frac{\sigma_{\rm DP}} {\sigma_{\rm hard}} = 
    \frac{\sigma^{\gamma j}}{\sigma_{\rm eff}} \frac{\sigma^{jj}}{\sigma_{\rm hard}},
    \end{equation} 
    where we used Eq.~(\ref{eq:sigma_DPS}).
    Then the number of DP events, $N_{\rm DP}$, can be expressed from $P_{\rm DP}$
    with a correction for the geometric and kinematic selection efficiency $\epsilon_{\rm DP}$,
    the single-vertex event selection efficiency $\epsilon_{\rm 1vtx}$, %fraction of DP events $f_{\rm DP}$ 
    and the number of beam crossings with one hard collision, $N_{\rm 1coll}$:
    \begin{eqnarray}    
    N_{\rm DP} = 
    ~\frac{\sigma^{\gamma j}}{\sigma_{\rm eff}} \frac{\sigma^{jj}}{\sigma_{\rm hard}}
    ~N_{\rm 1coll} ~\epsilon_{\rm DP} ~\epsilon_{\rm 1vtx}.
    \label{eq:ndp}
    \end{eqnarray}   

    The ratio of $N_{\rm DP}$ to $N_{\rm DI}$ allows us to obtain the expression for $\sigma_{\rm eff}$ in the following form: 
   \begin{eqnarray}   
   \sigma_{\rm eff} = \frac{N_{\rm DI}}{N_{\rm DP}} \frac{\ve_{\rm DP}}{\ve_{\rm DI}} R_{\rm c}\sigma_{\rm hard},
   \label{eq:sig_eff}
   \end{eqnarray}
   where $R_c \equiv (1/2) (N_{\rm 1coll} /N_{\rm 2coll}) (\ve_{\rm 1vtx} / \ve_{\rm 2vtx})$.
   The $\sigma^{\gamma j}$ and $\sigma^{jj}$ cross sections do not appear in this ratio
   and all the remaining efficiencies for DP and DI events enter only as ratios,
   resulting in a reduction of the impact of many correlated systematic uncertainties.

\begin{figure}[h]
\hspace*{-4mm} \includegraphics[scale=0.41]{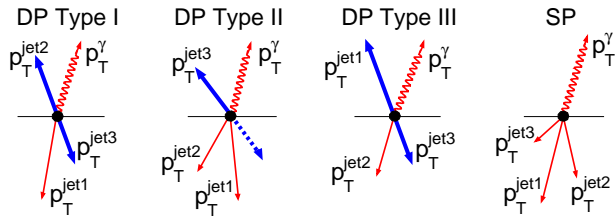}
~\\[-3mm]
\caption{Diagrams of DP Types I, II, III and SP \gpTHRj events.
For DP events, the light and bold lines correspond to two separate
parton interactions. The dotted line represents unreconstructed jet. }
\label{fig:dp_types}
\end{figure}

   Figure~\ref{fig:dp_types} shows the possible configurations of signal \gpTHRj DP events
   produced in a single  $p\bar{p}$  interaction and having one parton scattering in
   the final state with a $\gamma$ and at least one jet, superimposed with another parton scattering
   into a final state with at least one jet.
   We  define different event topologies as follows.
   Events in which both jets from the second parton scattering are reconstructed, 
   pass the selection cuts and are selected as the second and third jets, in
   order of decreasing jet $p_T$, are defined as Type~I.
   In Type II events, the second jet in the dijet process is either lost due to the finite
   jet reconstruction efficiency of detector acceptance or takes the fourth position after the jet $p_T$ ordering. 
   We also distinguish Type III events, 
   in which a jet from the second parton interaction 
   becomes the leading jet of the final 3-jets system, although they are quite rare
   given the $p_T$ range selected for the photon.

   The main background for the DP events are single parton (SP) scatterings with hard gluon bremsstrahlung
   in the initial or final state $qg \to q\gamma gg$, $q\bar{q} \to g\gamma gg$ that give the same
   \gpTHRj signature. They are also shown in Fig.~\ref{fig:dp_types}. 
   The fraction of DP events is determined in this analysis using a set of variables sensitive to the kinematic
   configurations of the two independent scatterings of parton pairs (see Secs. \ref{Sec:Vars} 
   and \ref{Sec:Frac}).

\begin{figure}[h]
~\\[2mm]
\hspace*{-4mm} \includegraphics[scale=0.41]{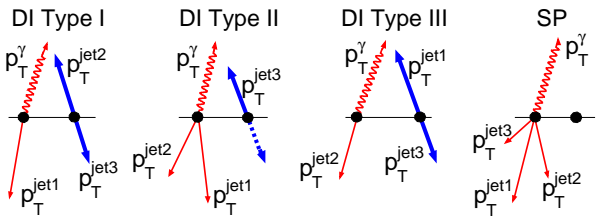}
~\\[-3mm]
\caption{Diagrams of DI Types I, II, III and SP \gpTHRj events.
For DI events, the light and bold lines correspond to two separate
$p\bar{p}$ interactions. The dotted line represents unreconstructed jet.}
\label{fig:di_types}
\end{figure}

   The DI events differ from the DP events by the fact that the second parton scattering happens
   at a separate $p\bar{p}$ collision vertex.
   The DI events, with the photon and at least one jet from one  $p\bar{p}$ collision,
   and at least one jet from another $p\bar{p}$ collision are shown in Fig.~\ref{fig:di_types}
   with a similar (to DP) set of DI event types. The background to DI events
   is due to two-vertex SP events with hard \gpTHRj events from one $p\bar{p}$ interaction
   with an additional soft interaction, i.e. having no reconstructed jets. The diagrams
   for these non-DI events are also shown in Fig.~\ref{fig:di_types}.


\section{D0 detector and data samples}
\label{Sec:ObjectID}

The D0 detector is described in detail in~\cite{D0_det}. Photon candidates are identified 
as isolated clusters of 
energy depositions in the uranium and liquid-argon sampling calorimeter. 
The central calorimeter covers the pseudorapidity~\cite{etaphi} range
$|\eta|<1.1$ and two end calorimeters
cover $1.5<|\eta|<4.2$.
The electromagnetic (EM) section of the calorimeter is segmented longitudinally into 
four layers and transversely into cells in pseudorapidity and azimuthal angle 
$\Delta\eta\times\Delta\phi=0.1 \times 0.1$ 
($0.05 \times 0.05$ in the third layer of the EM calorimeter).
The hadronic portion of the calorimeter is located behind the EM section.
The calorimeter surrounds 
a tracking system consisting of silicon microstrip and
scintillating fiber trackers, both located within a 2~T solenoidal magnetic field.

The events used in this analysis should first pass triggers based on the identification
of high $p_T$ clusters in the EM calorimeter with loose
shower shape requirements for photons.
These triggers are $100\%$ efficient for $\Ptg \gt 35$~GeV.
To select photon candidates in our data samples,
we use the following criteria~\cite{gamjet_PLB}.
EM objects are reconstructed using a simple cone algorithm with
a cone size ${\cal R}=\sqrt{(\Delta\eta)^2 + (\Delta\phi)^2}=0.2$.
Regions with poor photon identification capability and limited $\Ptg$ resolution,
at the boundaries between calorimeter modules and between the central and endcap 
calorimeters, are excluded from analysis.
Each photon candidate was required to deposit more than 96\% of detected energy
in the EM section of the calorimeter 
and to be isolated in the angular region between
${\cal R}=0.2$ and ${\cal R}=0.4$ around the 
center of the cluster:
$(E^{\rm iso}_{\rm Tot}-E^{\rm iso}_{\rm Core})/E^{\rm iso}_{\rm Core} < 0.07$, where $E^{\rm iso}_{\rm Tot}$ 
is overall (EM+hadronic) tower energy in the ($\eta,\phi$) cone of radius ${\cal R}=0.4$
and $E^{\rm iso}_{\rm Core}$ is EM tower energy within a radius of ${\cal R}=0.2$.
Candidate EM clusters matched to a reconstructed track are excluded from the
analysis. Clusters are matched to a reconstructed track by computing a $\chi^2$ function
which evaluates the consistency, within uncertainties, between the reconstructed
$\eta$ and $\phi$ positions of the cluster and of the closest track extrapolated to the finely-segmented
third layer of the EM calorimeter. The corresponding $\chi^2$
probability is required to be $<0.1\%$.
We also require the energy-weighted EM cluster width in the finely-segmented third EM layer
to be consistent with that expected for an electromagnetic shower.
In addition to the calorimeter isolation, we also apply a track isolation cut,
requiring the scalar sum of track transverse
momenta in a annulus of $0.05 \leq {\cal R} \leq 0.4$ to be less than 1.5~GeV.
Jets are reconstructed using 
the iterative midpoint cone algorithm~\cite{Run2Cone}
with a cone size of $0.7$. Jets must satisfy
quality criteria which suppress background from leptons, photons, and
detector noise effects.
To reject background from cosmic rays and $W\to\ell\nu$ decay,
the missing transverse momentum in the event is required to be less than $0.7p_{T}^{\gamma}$.
All pairs of objects in the event, (photon, jet) or (jet, jet), also are required to be
separated in $\eta-\phi$ space by $\Delta{\cal{R}}>0.7$.

Each event must contain
at least one $\gamma$ in the rapidity
region $|y|<1.0$ or $1.5<|y|<2.5$ 
and at least three jets with $|y|<3.0$.
Events are selected with $\gamma$ transverse momentum 
$60<p^{\gamma}_{T}<80$~GeV, leading (in $p_T$) jet $p_T>25$~GeV, while
the next-to-leading (second) and third jets must have $p_T>15$~GeV.
The jet transverse momenta are corrected to the particle level.
The high $p^{\gamma}_{T}$ scale
(i.e. the scale of the first parton interaction) 
allows a better separation of the first and second parton interactions in momentum space.

Data events with a single $p\bar{p}$ collision vertex, which compose the sample of DP candidates 
(``1Vtx'' sample), are selected separately from events with two vertices which compose 
the sample of DI candidates (``2Vtx'' sample).
The collision vertices in both samples
are required to have at least three associated tracks 
and to be within 60~cm of the center of the detector along the beam ($z$) axis. 

The $p_T$ spectrum for jets from dijet events falls faster than that for %radiation jets,
jets resulting from initial or final state radiation in the \gpj events, 
and thus DP fractions should depend on the jet $p_T$~\cite{Landsh,TH1,TH2,Sjost}.
The DP fractions and $\sigma_{\rm eff}$ are determined
in three $\ptsj$ bins: 
15--20, 20--25, and 25--30~GeV.
The total numbers of 1Vtx and 2Vtx \gpTHRj events remaining in each of the three $\ptsj$ bins 
after all selection criteria 
are given in Table~\ref{tab:12vtx_data}.
~\\[-7mm]
\begin{table}[htpb]
\begin{center}
\caption{The numbers of selected 1Vtx and 2Vtx \gpTHRj events in bins of $\ptsj$.}
\label{tab:12vtx_data}
\begin{tabular}{cccc} \hline\hline
  Data     & \multicolumn{3}{c}{ $\ptsj$ (GeV)} \\\cline{2-4}
  Sample & ~$15-20$~ & ~$20-25$~ & ~$25-30$ \\\hline
  1Vtx  ~& 2182 ~&  3475 ~&  3220 \\
  2Vtx  ~& 2026 ~&  2792 ~&  2309 \\\hline\hline
\end{tabular}
\end{center}
\end{table}

\section{DP and DI models}
\label{Sec:Models}
\begin{figure}[h]
\hspace*{-2mm} \includegraphics[scale=0.24]{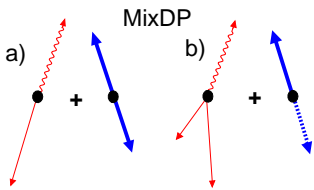}
~\\[-3mm]
\caption{Description of the mixing procedure used to prepare the MixDP signal sample.
Two combinations of mixing $\gamma+1$ jet and two jets from dijet events (a) and 
$\gamma+2$ jets and one jet from dijet events (b) are considered. 
The dotted line represents a jet failing the selection requirements.}
\label{fig:dp_mix}
\end{figure}

To study properties of DP and DI events and calculate their fractions in 
the 1Vtx and 2Vtx samples, respectively, we construct DP and DI models
by pairing data events. 
The DP model is constructed by overlaying 
in a single event one event of an inclusive sample of $\gamma+\!\geq$1 jet  events and 
one event of a sample of 
inelastic non-diffractive events selected with the minimum bias trigger and
a requirement of at least one jet (``MB'' sample)~\cite{MB}. 
Both samples contain only single-vertex events.
The jet $p_T$ from the MB events is recalculated relative to the vertex of  the $\gamma+$jet event.
The resulting mixed events, with jets 
re-ordered in $p_T$, are required to pass the \gpTHRj event selections described above.
This model of DP events, called {MixDP}, assumes
independent parton scatterings, with \gpj and dijet final states, by construction. 
The mixing procedure is shown schematically in Fig.~\ref{fig:dp_mix}.

In the DI model,  called {MixDI},
each event is constructed by mixing one event of the $\gamma+\!\geq$1 jet sample
and one event of the $\geq$1 jet MB sample.
Both events are exclusively selected from the two-vertices
events sample.
In the case of $\geq\!2$ jets in any 
component of the MixDI mixture, 
the first two jets, leading in $p_T$, are required to originate from the same vertex using 
the position along the beam axis of the point of closest approach to a vertex for the tracks 
associated to each jet and a cut on the minimal jet charged particle fraction, as discussed in Appendix B.
We consider the two-vertex \gpj and dijet events, components of the MixDI model, to better take into account
the underlying energy, coming from the soft interactions of the spectator partons.
The amount of this energy is different for single-- and two-vertex events and causes
a difference in the photon and jet identification efficiencies in the DP and DI events (see Section VII).
As a background to the DI events, we consider the two-vertex \gpTHRj sample
without a hard interaction at the second vertex ({Bkg2Vtx} sample),
obtained by imposing the direct requirement that all
three jets originate from the same vertex using the jet track information.

The fractions of Type~I~(II) events in the MixDP and MixDI samples
are the same within $1.5\%$ for each $\ptsj$ bin and 
vary for both samples  
from 26\%~(73\%) at $15\lt\ptsj\lt20$~GeV to (14--15)\%~[(84--86)\%] at $25\lt\ptsj\lt30$~GeV.
Type III events are quite rare and their fraction does not exceed $1\%$.
The {MixDP} and {MixDI} samples 
have similar kinematic ($p_T$ and $\eta$) distributions for the photon and all the jets.
They differ only
by the amount of energy coming from soft parton interactions
in either one or two $p\bar{p}$ collisions,
which may affect the photon and the jet selection efficiencies.

\section{Discriminating variables}
\label{Sec:Vars}

A distinctive feature of the DP events is the presence of two independent parton-parton 
scatterings within the same $p\bar{p}$ collision.
We define variables sensitive to the kinematics of DP events, specifically 
to the difference between the $p_T$ imbalance of the two object pairs in DP and SP \gpTHRj events as~\cite{TH2}: 
\begin{eqnarray}
\Delta S \equiv \Delta\phi\left(\vec{p}_{T}(\gamma,i), ~\vec{p}_{T}(j,k)\right),
\label{eq:Sphi_set}
\end{eqnarray}
where the indices $i, j, k ~(=1,2,3)$ run over the jets in the event.
Here $\vec{p}_{T}(\gamma,i) = {\vec p}_T^{~\gamma} + {\vec p}_T^{\rm ~jet_i}$ 
and $\vec{p}_{T}(j,k) = {\vec p}_T^{\rm ~jet_j} + {\vec p}_T^{\rm ~jet_k}$,
where the two object pairs, $(\gamma$, jet $i$) and (jet $j$, jet $k$),
are selected to give the minimal $p_T$ imbalance. These pairs are found by minimizing
$S_{p{_T}}$, or $S_{p^{\prime}_T}$, or $S_{\phi}$ defined as
\begin{eqnarray}
S_{p_{T}}=\frac{1}{\sqrt{2}}
\sqrt{{\left(\frac{\left|\vec{p}_{T}(\gamma,i)\right|}{\delta p_{T}(\gamma,i)}\right)}^{2}+
{\left(\frac{\left|\vec{p}_{T}(j,k)\right|}{\delta p_{T}(j,k)}\right)}^{2}},
\label{eq:S_var}
\end{eqnarray}
~\\[-4mm]
\begin{eqnarray}
S_{p^{\prime}_{T}}=\frac{1}{\sqrt{2}}
\sqrt{{\left(\frac{\left|\vec{p}_{T}(\gamma,i)\right|}{\left|\vec{p}_{T}^{~\gamma}\right| + \left|\vec{p}_{T}^{~i}\right|}\right)}^{2}
+ {\left(\frac{\left|\vec{p}_{T}(j,k)\right|}{\left|\vec{p}_{T}^{~j}\right| + \left|\vec{p}_{T}^{~k}\right|}\right)}^{2}},
\label{eq:Ss_var}
\end{eqnarray}
~\\[-4mm]
\begin{eqnarray}
S_{\phi}=\frac{1}{\sqrt{2}}
\sqrt{{\left[\frac{\Delta\phi(\gamma,i)}{\delta\phi(\gamma,i)}\right]}^{2}
+ {\left[\frac{\Delta\phi(j,k)}{\delta\phi(j,k)}\right]}^{2}}.
\label{eq:Sphi_var}
\end{eqnarray}

%In Eq.~(\ref{eq:Sphi_var}) $\Delta\phi(\gamma,i)$ is the supplement to $\pi$ of the azimuthal
%angle between the vectors ${\vec p}_{T}^{~\gamma}$ and ${\vec p}_{T}^{\rm ~jet_i}$.
%In Eq.~(\ref{eq:Sphi_var}) $\Delta\phi(\gamma,i)$  is the supplement to $\pi$ of 
%the minimal azimuthal angle between the vectors ${\vec p}_{T}^{~\gamma}$ and 
%${\vec p}_{T}^{\rm ~jet_i}$, $\phi(\gamma,i)$, i.e. it is defined as
%$\Delta\phi(\gamma,i) = |\pi - \phi(\gamma,i)|$.
In Eq.~(\ref{eq:Sphi_var}) $\Delta\phi(\gamma,i)= |\pi - \phi(\gamma,i)|$ is the supplement to $\pi$ of
the minimal azimuthal angle between the vectors ${\vec p}_{T}^{~\gamma}$ and
${\vec p}_{T}^{\rm ~jet_i}$, $\phi(\gamma,i)$.

The uncertainties $\delta p_{T}(\gamma,i)$ 
in Eq.~(\ref{eq:S_var}) and $\delta\phi(\gamma,i)$ 
in Eq.~(\ref{eq:Sphi_var})
are calculated as root-mean-square values of the $|\vec{p}_{T}(\gamma,i)|$
and $\Delta\phi(\gamma,i)$ distributions using the signal {MixDP} sample for each of the three possible pairings.
Azimuthal angles and uncertainties for jets $j$ and $k$
are defined analogously to those for the photon and jet $i$.
Any of the S-variables in Eqs.~(\ref{eq:S_var})--(\ref{eq:Sphi_var}) represents a significance of 
the pairwise $p_T$-imbalance.
On average, it should be higher for the SP events than for the DP events.
Also, each S-variable effectively splits the \gpTHRj system into $\gamma$+jet and dijet pairs,
based on the best pairwise balance.

The two best $p_T$-balancing pairs, which give the minimum $S$ for each of three variables 
in Eqs.~(\ref{eq:S_var}) -- (\ref{eq:Sphi_var}), are used to calculate 
the corresponding $\Delta S$ variables, $\Delta S_{p_{T}}, \Delta S_{p^{\prime}_{T}}$ and 
$\Delta S_{\phi}$, according to Eq.~(\ref{eq:Sphi_set}).
The $\Delta S_{p_{T}}, \Delta S_{p^{\prime}_{T}}$ variables are also used in~\cite{UA2,CDF97}, while
the $\Delta S_{\phi}$ is first introduced in this measurement. 

\begin{figure}[h]
~\\[-4mm]
\hspace*{2mm} \includegraphics[scale=0.35]{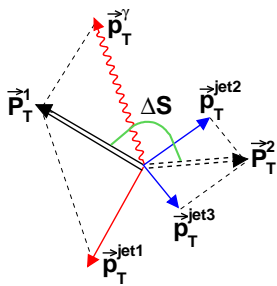}
~\\[-9mm]
\caption{A possible orientation of photon and jets transverse momenta vectors in \gpTHRj events.
Vectors $\vec{\rm P}_{\rm T}^{1}$ and $\vec{\rm P}_{\rm T}^{2}$ are the $p_T$ imbalance vectors 
of $\gamma$+jet and jet-jet pairs. 
The figure illustrates a general case for the production of \gpTHRj+X events.}
\label{fig:sp}
~\\[-9mm]
\end{figure}

Figure~\ref{fig:sp} illustrates a possible orientation of the transverse momenta vectors of the 
photon and jets as well as their $p_T$ imbalances vectors, $\vec{\rm P}_{\rm T}^{1}$ and $\vec{\rm P}_{\rm T}^{2}$, 
in \gpTHRj events. 
In SP events, the topologies with the two radiation jets emitted close to the leading jet
(recoiling against the photon direction in $\phi$) are preferred. 
The resulting peak at $\Delta S = \pi$ is smeared by the effects of 
additional gluon radiation and detector resolution.
For a simple model of DP events, we have exact pairwise balance in $p_T$ and thus
$\Delta S$ will be undefined. The exact $p_T$ balance in the pairs can be violated 
due to either detector resolution or additional gluon radiation. 
Both effects introduce an additional random contribution
to the azimuthal angle between the $\gamma$+jet and the
dijet $p_T$ imbalance vectors, broadening the $\Delta S$
distribution (see also Fig.~\ref{fig:DeltaS_tunes} below).

\section{Fractions of DP and DI events}
\label{Sec:Frac}

\subsection{Fractions of DP events}

\begin{figure}[h]
\vspace*{-2mm}
\hspace*{4mm} \includegraphics[scale=0.30]{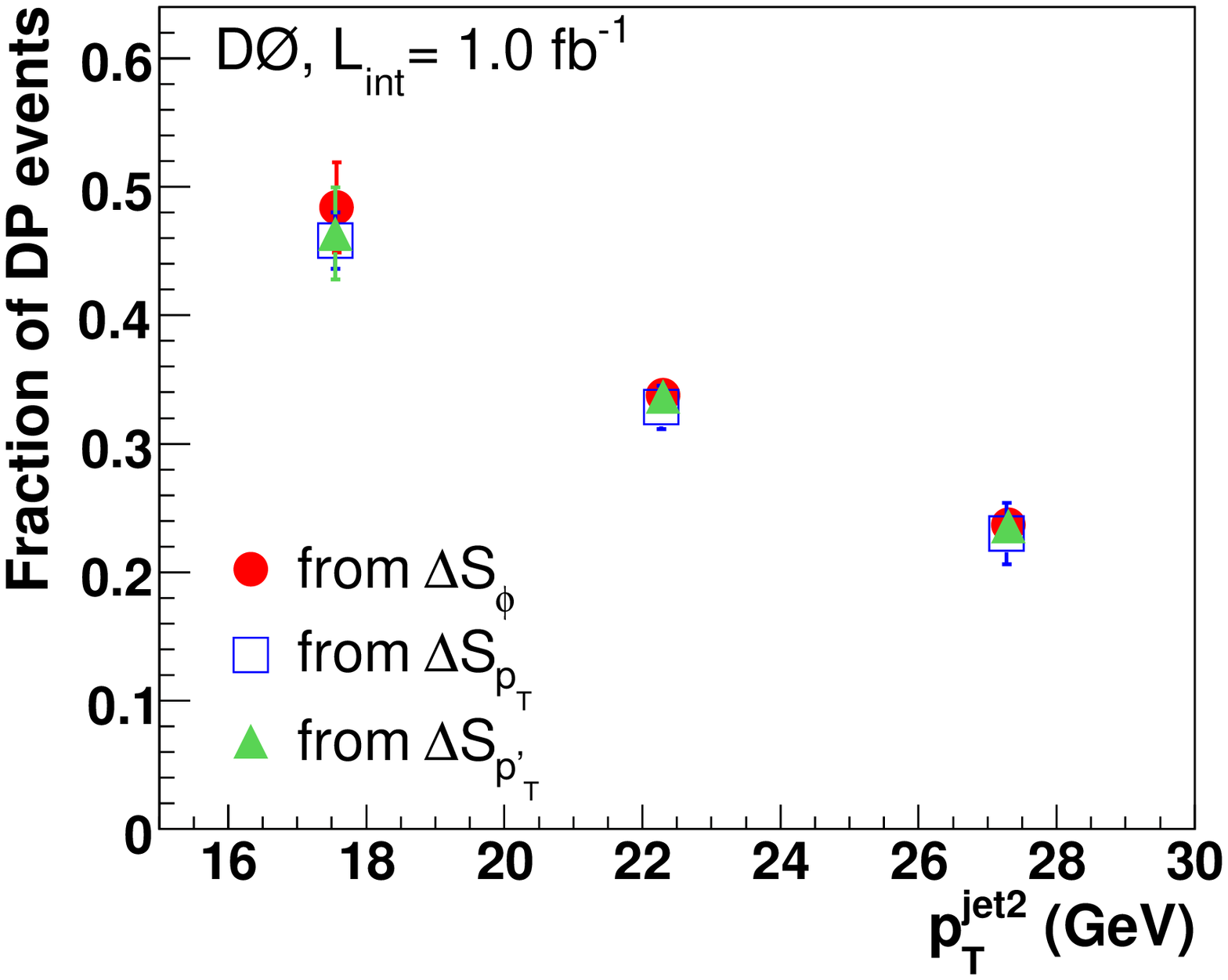}
\vspace*{-3mm}
\caption{Fractions of DP events extracted with the \dSphi, \dSpt, and \dSptp variables in the three $\ptsj$ intervals.}
\label{fig:dp_frac}
\end{figure}

To extract the fractions of  DP events, we 
exploit the difference in the $p_T$ spectrum of DP and radiation jets, mentioned  
in Sec. \ref{Sec:ObjectID}, and consider data in
two adjacent $\ptsj$ intervals: DP-enriched at smaller $\ptsj$ and DP-depleted  
at larger $\ptsj$~\cite{Landsh,TH1,TH2}.
The distribution for each $\Delta S$  variable in data ($D$) can be expressed
as a sum of signal (DP) and background (SP) distributions:\\[-5mm]
\begin{eqnarray}
D_1=f_1M_1+(1-f_1)B_1
\label{eq:d1}
\end{eqnarray}
~\\[-10mm]
\begin{eqnarray}
D_2=f_2M_2+(1-f_2)B_2,
\label{eq:d2}
\end{eqnarray}
where $M_i$ and $B_i$ stand for the signal MixDP and background distributions,
$f_i$ is the DP fraction, $(1-f_i)$ is the SP fraction,
and indices 1, 2 correspond to the DP-enriched and DP-depleted data sets.
Multiplying (\ref{eq:d2}) by $\lambda K$ and subtracting from (\ref{eq:d1})
we obtain:
\begin{eqnarray}
D_1-\lambda KD_2 = f_1M_1-\lambda KCf_1M_2,
\label{eq:syst3}
\end{eqnarray}
where $\lambda=B_1/B_2$ is the ratio of the background distributions, and
$K=(1-f_1)/(1-f_2)$ and $C=f_2/f_1$ are the ratios
of the SP and DP fractions between the DP-enriched and DP-depleted samples, respectively.
In contrast to~\cite{CDF97}, we introduce a factor $\lambda$ that 
corrects for the relative difference of $\Delta S$ shapes for the SP distributions
in adjacent $\ptsj$ intervals.
It is obtained using Monte Carlo (MC) \gpTHRj events generated with {\sc pythia}~\cite{PYT}
without multiple parton interactions
and with a full simulation of the detector response and 
is found to be in the range $0.95-1.3$ for different bins of $\Delta S$.
The factor $C$ is extracted
using ratios of the numbers of events
in data and MixDP samples in the adjacent bins by
\begin{eqnarray}
C = \left({N_{2}^{\rm MixDP}}/{N_{2}^{\rm data}}\right)/\left({N_{1}^{\rm MixDP}}/{N_{1}^{\rm data}}\right),
\label{eq:C1}
\end{eqnarray}
i.e. without actual knowledge of DP fractions in those bins.
Thus, the only unknown parameter in Eq.~(\ref{eq:syst3}) is the DP fraction $f_1$.
It is obtained from a $\chi^2$ minimization of Eq.~(\ref{eq:syst3})
using {\sc minuit}~\cite{Minuit}.
The fit was performed for each pair
of $\ptsj$ bins ($15-20/20-25$~GeV and $20-25/25-30$~GeV) and for each
of $\Delta S$ variables (\ref{eq:S_var})--(\ref{eq:Sphi_var}). 
The DP fractions in the last bin, $25 < \ptsj < 30$~GeV, are calculated from $f_2=Cf_1$.
The extracted DP fractions are shown in Fig.~\ref{fig:dp_frac}.
The DP fractions, averaged over the three $\Delta S$ variables (with uncertainties),
are summarized in Table~\ref{tab:dp_frac}.
The location of the points in Fig.~\ref{fig:dp_frac} corresponds
to the mean $\ptsj$ for the DP model in a given bin. They are also
shown in Table~\ref{tab:dp_frac} as $\la \ptsj \ra$.
The uncertainties are mainly caused by the statistics of the data and MixDP samples (used in the fitting) 
and partially by the determination of~$\lambda$ $(2-5)\%$.
%~\\[-5mm]
\begin{table}[htpb]
\small
\caption{Fractions of DP events in the three $\ptsj$ bins.}
\label{tab:dp_frac}
\begin{tabular}{ccc} \hline\hline
$\ptsj$ GeV ~&~ $\la \ptsj \ra$ (GeV) ~&~ $f_{\rm DP}$ \\\hline
$15-20$ ~&~ $17.6$ ~&~ $0.466 \pm 0.041$ \\
$20-25$ ~&~ $22.3$ ~&~ $0.334 \pm 0.023$ \\
$25-30$ ~&~ $27.3$ ~&~ $0.235 \pm 0.027$ \\\hline\hline
\end{tabular}
\end{table}
%~\\[-5mm]

Since each component of a MixDP signal event may contain
two jets, where one jet may be caused by an additional parton interaction, the MixDP sample
should simulate the properties of the double plus triple parton (TP) interactions (DP+TP), and thus the fractions in 
Table~\ref{tab:dp_frac} take into account a contribution from triple interactions as well.
In this sense, the DP cross section %$\sigma_{\rm eff}$ that we are obtaining in Section \ref{Sec:Sigma_eff},
calculated using Eqs.~(\ref{eq:sigma_DPS}) and (\ref{eq:sig_eff}) 
is inclusive~\cite{CDF_excl,Treleani_2007}.

\begin{figure}[t]
\hspace*{-2mm} \includegraphics[scale=0.43]{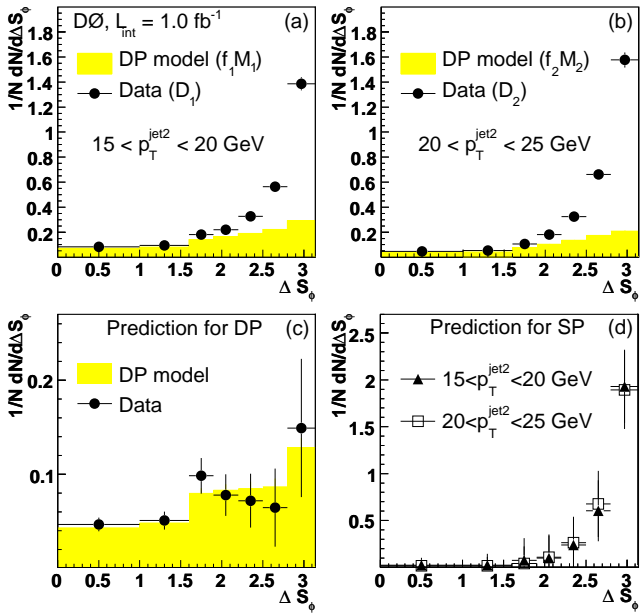}
~\\[-1mm]
\caption{Results of the two datasets fit for the $\Delta S_{\phi}$ variable
for the combination of two $\ptsj$ bins $15-20$~GeV and $20-25$~GeV. (a) and (b)
show distributions for data (points) and the DP model (shaded area);
(c) shows the prediction for DP from data (points), corrected to remove SP contribution, and
the DP model (shaded area) as a difference between the corresponding distributions of (a) and (b);
(d) shows the extracted SP distributions in the two bins.
The error bars in (a) and (b) are only statistical, while
in (c) and (d) they represent total (statistic and systematic) uncertainty.
}
\label{fig:fit_test1}
\vskip -3mm
\end{figure}
\begin{figure}[t]
\hspace*{-2mm} \includegraphics[scale=0.43]{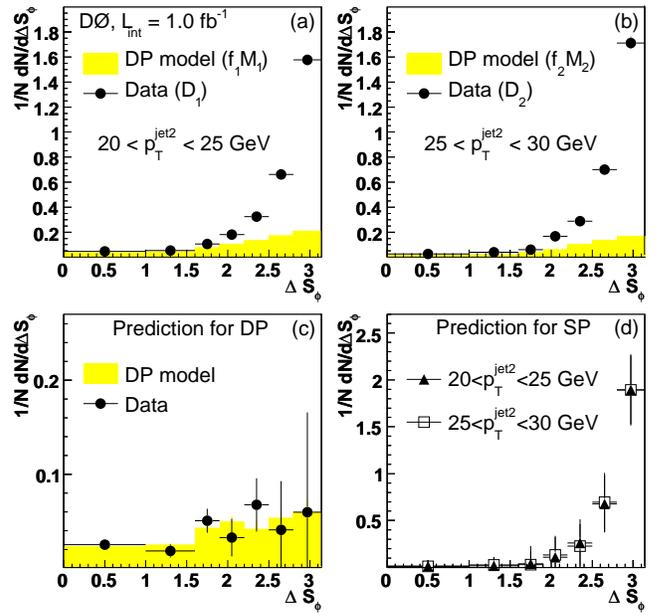}
~\\[-1mm]
\caption{Same as in Fig.~\ref{fig:fit_test1} but for
the other combination of $\ptsj$ bins, $20-25$~GeV and $25-30$~GeV.}
\label{fig:fit_test2}
\vskip -3mm
\end{figure}

Figure \ref{fig:fit_test1} shows tests of the fit results for $f_1$
using the $\Delta S_{\phi}$ variable for the combination of two $\ptsj$ bins, $15-20$~GeV 
and $20-25$~GeV.
Figure~\ref{fig:fit_test1}(a) show the $\Delta S_{\phi}$ distributions for the DP-enriched data set
in data ($D_1$) and the {MixDP} sample ($M_1$) weighted with its fraction $f_1$.
Figure~\ref{fig:fit_test1}(b) shows analogous distributions for the DP-depleted dataset: data ($D_2$)
and the {MixDP} sample ($M_2$) weighted with its fraction $f_2$.
It can be concluded from the two distributions
that the regions of small $\Delta S_{\phi}$ ($\lesssim 1.5$)
is mostly populated by signal 
events with two independent hard interactions.
Figure~\ref{fig:fit_test1}(c) shows the difference between the data distributions of
Figs.~\ref{fig:fit_test1}(a) and \ref{fig:fit_test1}(b), 
corrected to remove the SP contribution by the factor $\lambda K$
(the factor $\lambda$ corrects for the relative difference of the $\Delta S_{\phi}$ shapes 
and $K$ corrects for the difference in the SP fractions in the two adjacent $\ptsj$ bins)
[left side of Eq.~(\ref{eq:syst3})] and compared to
the MixDP prediction [right side of Eq.~(\ref{eq:syst3})].
As expected, the difference is always positive since the fractions of DP events drop with $\ptsj$.
The DP model provides an adequate description of the data.
In Fig.~\ref{fig:fit_test1}(d) we extract the SP distributions  
by subtracting the estimated DP contributions from the data:
$(D_1-f_1M_1)/(1-f_1)$
for the DP-enriched data set and $(D_2-f_2M_2)/(1-f_2)$
for the DP-depleted data sets.
Figure~\ref{fig:fit_test2} shows the analogous test of the fit results  
for the other pair of $\ptsj$ bins, $20-25$~GeV and $25-30$~GeV.

\begin{figure}[h]
~\\[-2mm]
\hspace*{5mm} \includegraphics[scale=0.26]{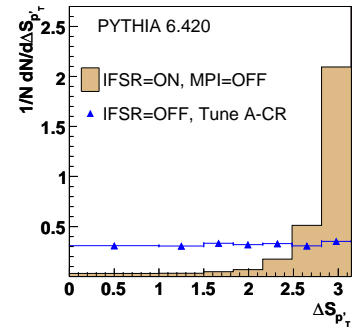}
\vskip -3mm
\caption{\dSptp distributions for \gpTHRj events simulated using {\sc pythia} with ISR/FSR but with MPI switched off
(shaded region), as well as for \gpTHRj events without ISR/FSR but MPI switched on using Tune A-CR 
(triangle markers). The bin $15<\ptsj<20$~GeV is considered.}
\label{fig:DeltaS_tunes}
\vskip -3mm
\end{figure}

Predictions for SP events are obtained using {\sc pythia}.
The \dSptp distribution for \gpTHRj events simulated with initial
and final state radiation (ISR and FSR) and without multiple parton interactions (MPI) 
is shown in Fig.~\ref{fig:DeltaS_tunes} for the interval $15<\ptsj<20$~GeV.
Since the $\vec{p}_{T}$ imbalance of the two additional jets
should compensate the $\vec{p}_{T}$ imbalance of the ``$\gamma$+leading jet'' system,
the \dSptp distribution is shifted towards $\pi$.
This distributions shows good agreement with the results for the SP sample shown in 
Fig.~\ref{fig:fit_test1}(d).
The DP \gpTHRj events are also simulated 
without ISR and FSR and using the MPI model corresponding to the {\sc pythia} 
parameters Tune A-CR~\cite{PYT}.
In this case, the two subleading jets 
may originate only from the second parton interaction 
(as in DP events of Type I, see Fig.~\ref{fig:dp_types}).
As expected, the \dSptp distribution for these events is uniform,
since the two $p_T$ balance vectors for the two systems, \gpj and dijets, 
are independent from each other. 

Another source of background to the single-vertex \gpTHRj DP events 
is caused by double $p\bar{p}$ collisions close to each other along the beam direction, 
for which a single vertex is reconstructed.
This was estimated separately and found to be negligible with a probability $<10^{-3}$.

\subsection{Fractions of DI events}

The DI fractions, $f_{\rm DI}$, 
are extracted by fitting the shapes of the $\Delta S$ distributions
of the {MixDI} signal and {Bkg2Vtx} background samples
to that for the 2Vtx data using the technique described in~\cite{HMCMLL}.
Uncertainties are mainly caused by the fitting procedure and by building Bkg2Vtx and MixDI 
(in case of Type I events) models.
To estimate the uncertainty due to the Bkg2Vtx or MixDI  models, 
we vary a cut on the minimal jet charged particle fraction (see Appendix B)
from 0.5 to 0.75. 
The fitted $f_{\rm DI}$ 
in this case varies in different $\ptsj$ bins within $(3-10)\%$,
which is taken as the uncertainty.
The final $f_{\rm DI}$ values with total uncertainties
are  $0.189 \pm 0.029$ for $15\!<\!\ptsj\!<\!20$~GeV,
$0.137\pm 0.027$ for $20\!<\!\ptsj\!<\!25$~GeV, and $0.094 \pm 0.025$ for $25\!<\!\ptsj\!<\!30$~GeV.
%The relative $f_{\rm DI}$ uncertainties grow with
%increasing $\ptsj$. This is caused by
%a decreasing probability for a jet to originate from a $p\bar{p}$ collision vertex
%additional to the main one, and, as a consequence, less sensitivity
%to such events in the 2Vtx data sample.
The relative $f_{\rm DI}$ uncertainties grow with
increasing $\ptsj$. This is caused by
a decreasing probability for a jet to originate from a second
$p\bar{p}$ collision vertex. As a consequence, the sensitivity
to DI events in the 2Vtx data sample becomes smaller.

\begin{figure}[h]
\hspace*{-7mm} \includegraphics[scale=0.23,bb=10 20 500 500,clip=true]{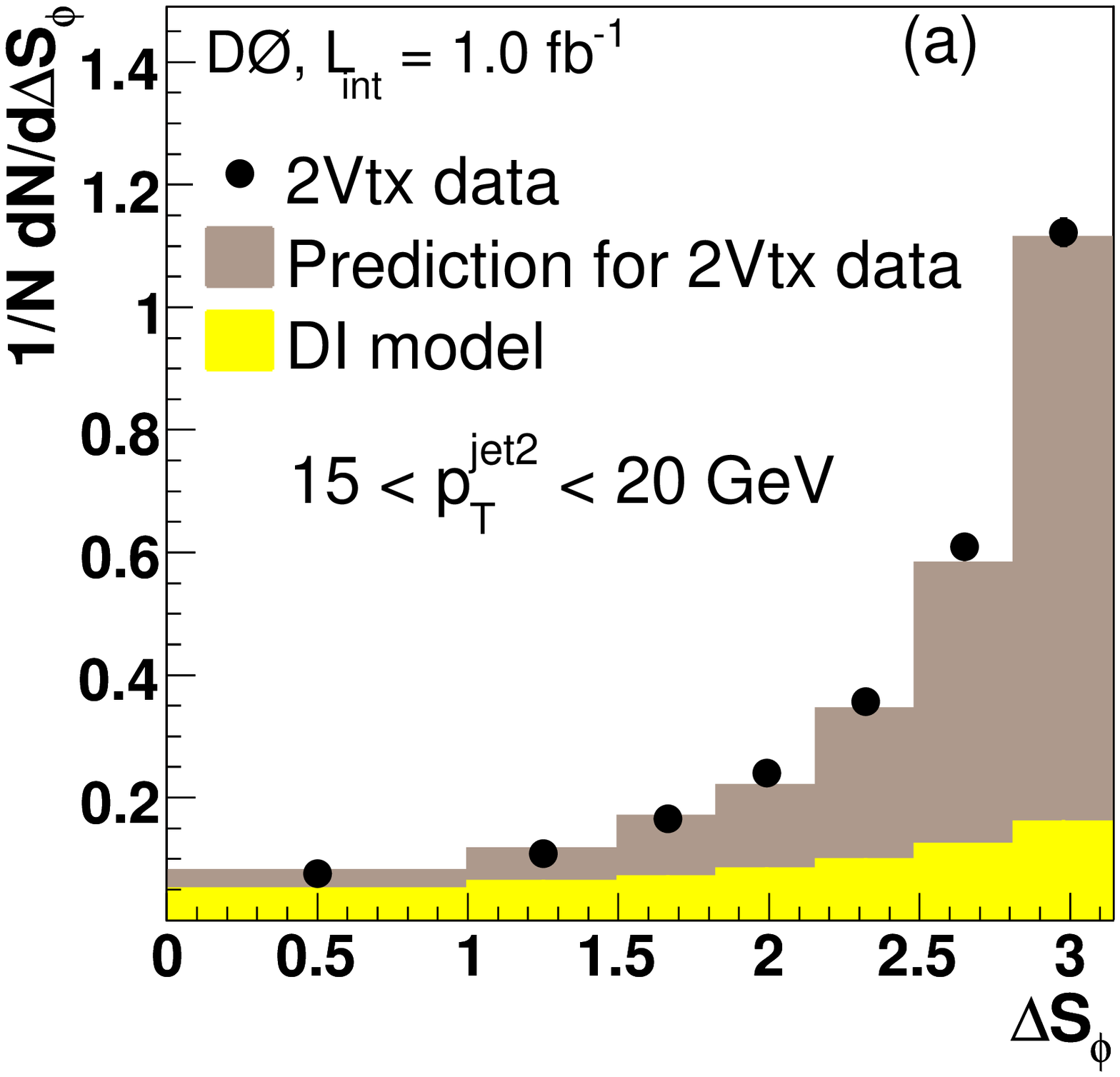}
\hspace*{1mm} \includegraphics[scale=0.23,bb=10 20 500 500,clip=true]{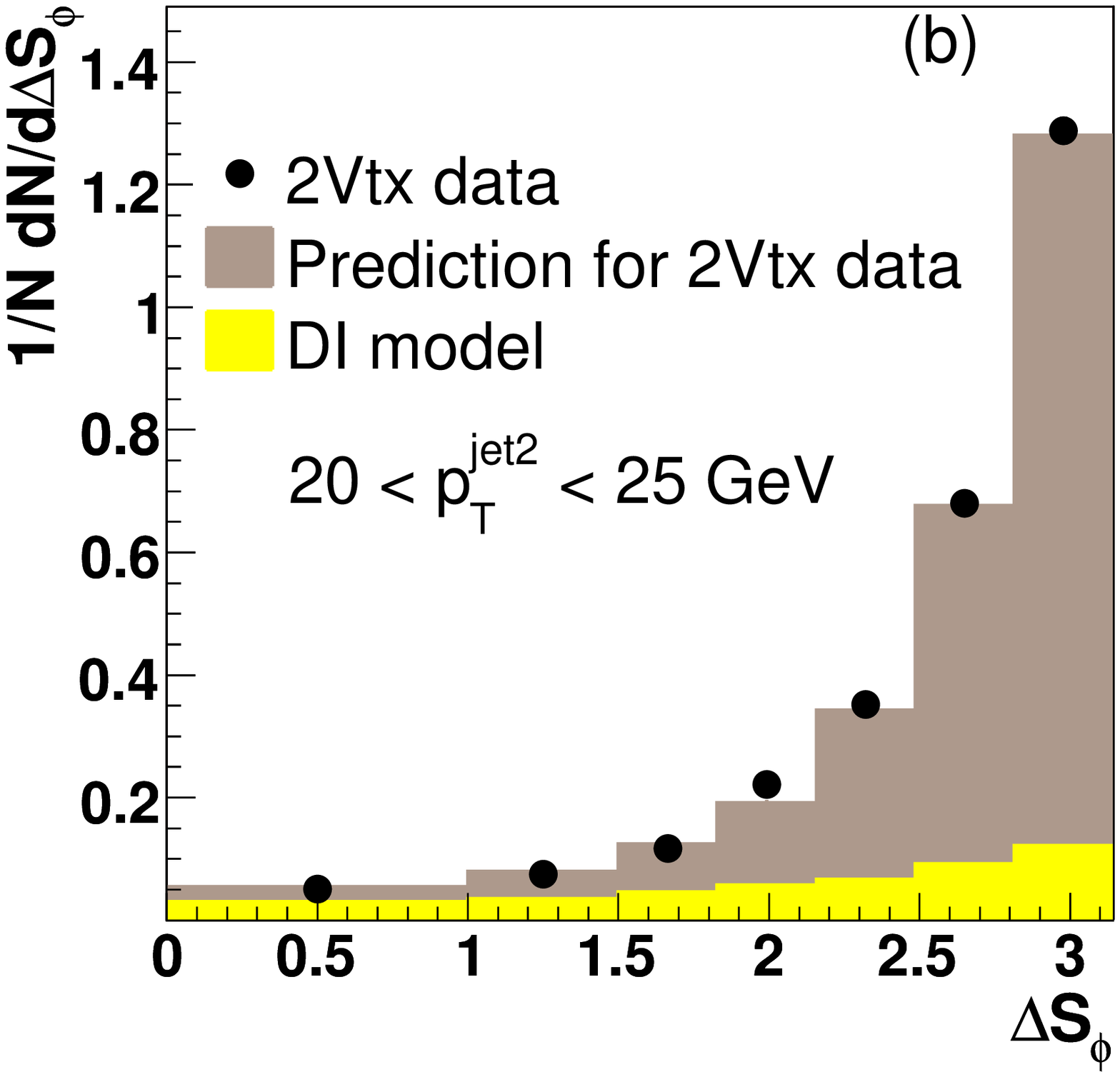}
\\[1mm]
\hspace*{-2mm} \includegraphics[scale=0.23,bb=10 20 500 500,clip=true]{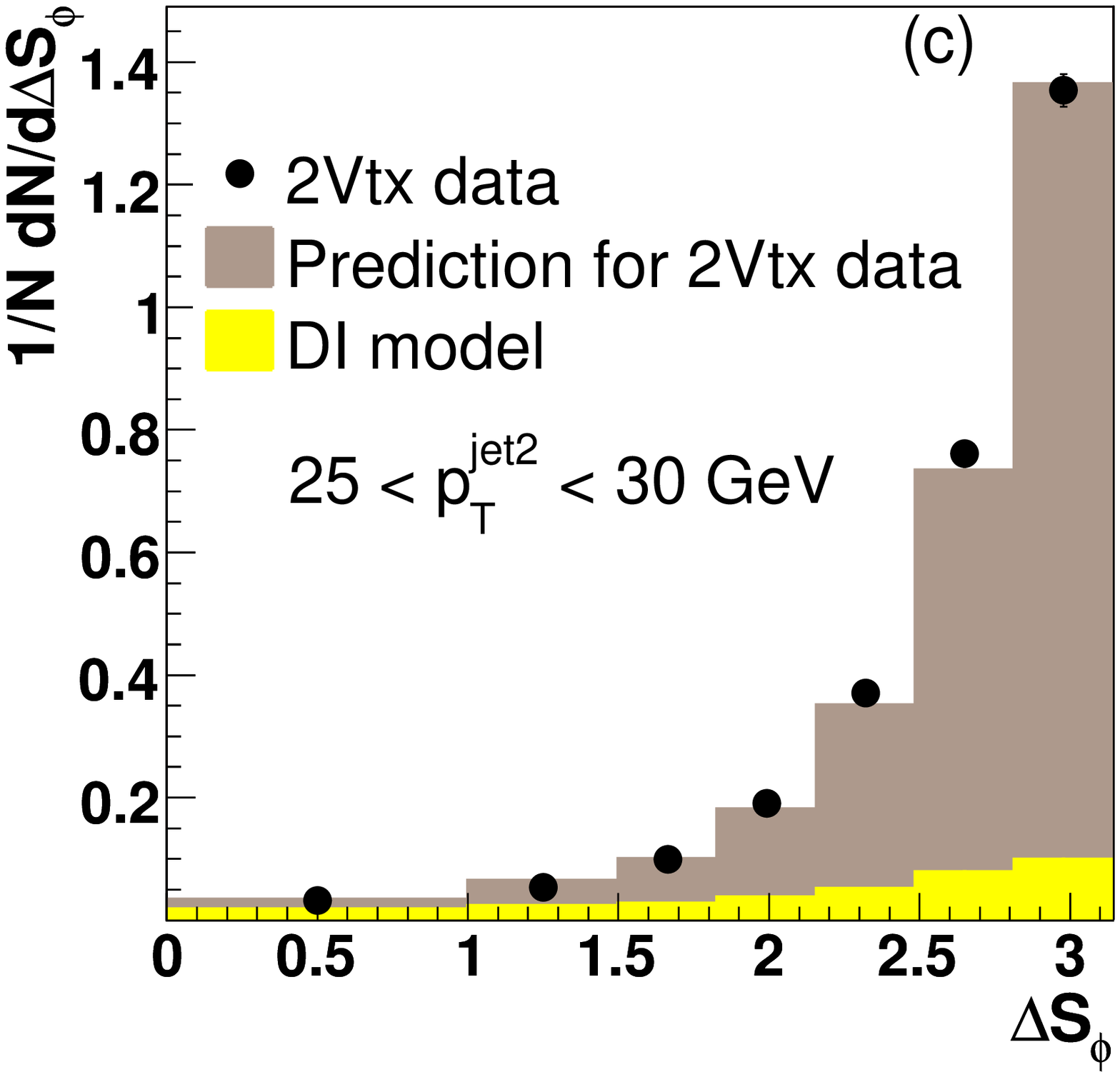}
\caption{$\Delta S_{\phi}$ distributions for two-vertex \gpTHRj events in the three $\ptsj$ intervals:
(a) $15-20$~GeV, (b) $20-25$~GeV and (c) $25-30$~GeV.  MixDI and the total sum of the MixDI and Bkg2Vtx distributions
(shaded histograms) are weighted with their fractions found from the fit, compared to 2Vtx data (black points).
The shown uncertainties are only statistical.
}
\label{fig:DeltaS_di}
\vskip -3mm
\end{figure}

Figure~\ref{fig:DeltaS_di} shows the $\Delta S_{\phi}$ distributions for the two-vertex \gpTHRj events 
selected in three $\ptsj$ intervals, $15-20$~GeV, $20-25$~GeV and $25-30$~GeV,
for the DI model (MixDI) and the total sum of MixDI and Bkg2Vtx distributions,
weighted with the DI fraction, and compared to 2Vtx data.
The weighted sums of the signal and background samples reproduce 
the shapes of the data distributions. 

\section{DP and DI efficiencies, $R_c$ and $\sigma_{\rm hard}$}
\label{Sec:Eff}

\subsection{Ratio of photon and jet efficiencies in DP and DI events}

The selection efficiencies for DP and DI events enter Eq.~(\ref{eq:sig_eff})
only as ratios, canceling many common correction factors and
correlated systematic uncertainties.
The DP and DI events differ from each other by the number of $p\bar{p}$ collision
vertices  (one vs. two), and therefore 
their selection efficiencies $\epsdi$ and $\epsdp$ may differ
due to different amounts of soft 
unclustered energy in the single and double
$p\bar{p}$ collision events.
This could lead to a difference in the jet reconstruction efficiencies,
due to the different probabilities of passing the jet selection requirement $p_T \gt 6$~GeV
(applied during jet reconstruction)
and different photon selection efficiencies, due to different amount of  
energy in the track and calorimeter isolation cones around the photon.

To estimate these efficiencies, we use  \gpj and dijet MC events
and also MixDI and MixDP data samples.
The  MC events are generated with {\sc pythia}~\cite{PYT} and processed
through a {\sc geant}-based~\cite{GEANT} simulation of the D0 detector 
response. In order to accurately model the effects of multiple proton-antiproton interactions
and detector noise, data events from random $p\bar{p}$ crossings are overlaid
on the MC events using data from the same time period as considered in the analysis.
These MC events are then processed using the same reconstruction code as for the data.
We also apply additional smearing to the reconstructed photon and jet $p_T$
so that the measurement resolutions in MC match those in data.
The MC events are preselected with the vertex cuts 
and split into the single-- and two-vertex samples.

The efficiencies for the photon selection criteria are estimated
using \gpj MC events. We found that the ratio
of photon efficiencies in single-vertex ($\varepsilon_{\rm 1v}^\gamma$) 
to that in two-vertex samples ($\varepsilon_{\rm 2v}^\gamma$) 
does not have a noticeable dependence on $\ptsj$ and can be taken as 
$\varepsilon_{\rm 1v}^\gamma /\varepsilon_{\rm 2v}^\gamma = 0.96\pm 0.03$. 
The purity of \gpj events in the interval of $60<\Ptg<80$~GeV in data is expected to be 
about $75\%$~\cite{gamjet_PLB}, and the remaining 
events are mostly dijet events with one jet misidentified as photon.
An analogous analysis of the MC dijet events gives the ratio of the efficiencies for jets
to be misidentified as photons equal to $0.99\pm 0.06$,
which does not change the $\varepsilon_{\rm 1v}^\gamma/\varepsilon_{\rm 2v}^\gamma$ value found with the signal \gpj sample.

The ratio of jet efficiencies is calculated in two steps.
First, the efficiencies
are estimated with respect to a requirement to have at least three jets with 
$p_T^{\rm jet1}\!>\!25$~GeV, $p_T^{\rm jet2}\!>\!15$~GeV, and $p_T^{\rm jet3}\!>\!15$~GeV.
These efficiencies are calculated using MC \gpj and dijet events 
mixed according to the fractions of the three main MixDP and MixDI event types, described in
Sec.~\ref{Sec:Models}.
The ratio of efficiencies for other jet selections 
(e.g. to get into the $\ptsj$ interval and satisfy $\Delta{\cal{R}}$ and jet rapidity selections)
has been calculated using MixDP and MixDI signal data samples.
The total ratio of DP/DI jet efficiencies is found to be stable for all $\ptsj$ bins and 
equal to $0.93$ with $\sim 5\%$ uncertainty. 
Thus, the overall ratio of photon and jet DP/DI selection efficiencies $\epsdp/\epsdi$ 
is about $0.90$ with uncertainties in the three $\ptsj$ bins varying within
$(5.6-6.5)\%$. 

\subsection{Vertex efficiencies}

The vertex efficiency $\varepsilon_{\rm 1vtx}$ ($\varepsilon_{\rm 2vtx}$) corrects
for the single (double) collision events that are lost
in the DP (DI) candidate sample due to the single (double) vertex cuts
($|z_{\rm vtx}| < 60~\text{cm}$ and $\geq 3$ tracks). 
The ratio $\varepsilon_{\rm 1vtx}/\varepsilon_{\rm 2vtx}$
is calculated from the data and found to be $1.08 \pm 0.01$
for all $\ptsj$ bins.
The probability to miss a hard interaction event with at least one jet with $p_T>15$~GeV
due to a non-reconstructed vertex is calculated in \gpj and minimum bias data and found to be $(0.2-0.4)\%$.
The probability to have an additional reconstructed vertex, passing the vertex selection requirements,
is estimated separately using \gpj and dijet MC events with at least one reconstructed jet with $p_T>15$~GeV 
and found to be less than $0.3\%$.

\subsection{Calculating $\sigma_{\rm hard}, N_{\rm 1coll}$ and $N_{\rm 2coll}$}

The numbers of expected events with one ($N_{\rm 1coll}$) and two ($N_{\rm 2coll}$) $p\bar{p}$ collisions
resulting in hard interactions are calculated from 
the known instantaneous luminosity  spectrum of the collected data ($L_{\rm inst}$), 
the frequency of beam crossings ($f_{\rm cross}$) for 
the Tevatron~\cite{D0_det},
and the hard $p\bar{p}$ interaction cross section ($\sigma_{\rm hard}$). 

The value of $\sigma_{\rm hard}$ at $\sqrt{s}=1.96$ TeV is obtained in the following way.
We use the inelastic cross section calculated at $\sqrt{s}=1.96$ TeV,
$\overline{\sigma}_{\rm inel}(1.96~{\rm TeV})=60.7 \pm 2.4 {~\rm mb}$~\cite{Klim}, found
from averaging  the inelastic cross sections measured by the CDF~\cite{CDF_xsec_tot} 
and E811~\cite{E811_xsec_tot} Collaboration at $\sqrt{s}=1.8$ TeV and extrapolated to 1.96 TeV.
To calculate  single diffractive (SD) and double diffractive (DD) cross sections
at $\sqrt{s}=1.96$ TeV, $\sigma_{\rm SD}(1.96~{\rm TeV})$ and $\sigma_{\rm DD}(1.96~{\rm TeV})$,
we use SD and DD cross sections measured at $\sqrt{s}=1.8$ TeV 
($\sigma_{\rm SD}(1.8~{\rm TeV})=9.46 \pm 0.44 {~\rm mb}$~\cite{CDF_xsec_tot}
and $\sigma_{\rm DD}(1.8~{\rm TeV})= 6.32 \pm 0.03({\rm stat}) \pm 1.7({\rm syst}){~\rm mb}$)~\cite{CDF_xsec-2Dif}
and extrapolate them to $\sqrt{s}=1.96$ TeV using the slow asymptotic behaviour predicted in~\cite{SCHUL}. 
We find \\[-5mm]
\begin{eqnarray}
\nonumber
\sigma_{\rm hard} (1.96~{\rm TeV}) = \overline{\sigma}_{\rm inel}(1.96~{\rm TeV}) - \sigma_{\rm SD}(1.96~{\rm TeV}) \\
- \sigma_{\rm DD}(1.96~{\rm TeV}) = 44.76 \pm  2.89 ~{\rm mb}.~
\label{eq:s_hard}
\end{eqnarray}
We also do analogous estimates by calculating first $\sigma_{\rm hard}$ at $\sqrt{s}= 1.8$ TeV
and then extrapolating it to $\sqrt{s}= 1.96$ TeV using~\cite{SCHUL}.
This method results in $\sigma_{\rm hard} (1.96~{\rm TeV}) = 43.85 \pm  2.63$~mb
which agrees well with Eq.~(\ref{eq:s_hard}).

In each bin of the  $L_{\rm inst}$ spectrum, we calculate 
the average number of hard $p\bar{p}$ interactions
$\la n \ra = (L_{\rm inst} / f_{\rm cross} ) \sigma_{\rm hard}$ 
and then $N_{\rm 1coll}$ and $N_{\rm 2coll}$ are determined from $\la n \ra$ using Poisson statistics. 
Summing over all $L_{\rm inst}$ bins, weighted with their fractions, 
we get $N_{\rm 1coll}/(2 N_{\rm 2coll}) = 1.169$ and thus $R_c \sigma_{\rm hard} = 56.45 \pm 0.88$ mb.
Here we take into account that $R_c$ and $\sigma_{\rm hard}$ enter Eq.~(\ref{eq:sig_eff})
for $\sigma_{\rm eff}$ as a product. Any increase of $\sigma_{\rm hard}$ leads to an increase of $\la n \ra$ and, 
as a consequence, to a decrease in $R_c$, and vice versa. 
Specifically, while the found value of $\sigma_{\rm hard}$ has a 6.5\% relative uncertainty, 
the product $R_c \sigma_{\rm hard}$ has approximately 2\% uncertainty.

\section{Results}
\label{Sec:Sigma_eff}

\subsection{Effective cross section}

\begin{figure}[h]
\vspace*{-2mm}
\hspace*{4mm} \includegraphics[scale=0.30]{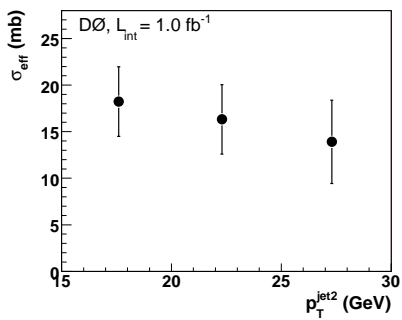}
~\\[-1mm]
\caption{Effective cross section $\sigma_{\rm eff}$ (mb) measured in the three $\ptsj$ intervals.}
\label{fig:sigma_eff}
\vskip -3mm
\end{figure}

The calculation of $\sigma_{\rm eff}$ is based on Eq.~(\ref{eq:sig_eff}) of Sec.~\ref{Sec:Intro}.
The numbers $N_{\rm DP}$ and $N_{\rm DI}$ in each $\ptsj$ bin are obtained from the numbers
of the 1Vtx and 2Vtx \gpTHRj events in Table~\ref{tab:12vtx_data},
multiplying them by $f_{\rm DP}$ and $f_{\rm DI}$.
The determination of all other components of Eq.~(\ref{eq:sig_eff}) are described in Sec.~\ref{Sec:Eff}.
The resulting values of $\sigma_{\rm eff}$ with total uncertainties
(statistical and systematic are summed in quadrature) 
 are shown in Fig.~\ref{fig:sigma_eff} and given in Table~\ref{tab:sigma_eff} for the three $\ptsj$ bins.
The location of the points in Fig.~\ref{fig:sigma_eff} corresponds
to the mean $\ptsj$ for the DP model in a given bin (the mean $\ptsj$ values for DI model are the same
within 0.15~GeV). These values are also shown in Table~\ref{tab:sigma_eff}.
Table~\ref{tab:uncert} summarizes the main sources of uncertainties for each $\ptsj$ bin. 
The main systematic uncertainties are related to the determinations of the DI fractions (dominant uncertainty),
DP fractions, the $\varepsilon_{\rm DP}/\varepsilon_{\rm DI}$ ratio, jet energy scale (JES), 
and $R_c \sigma_{\rm hard}$, giving a total systematic uncertainty of $(20.5-32.2)\%$. 
\begin{table}[htpb]
~\\[-3mm]
\small
\caption{Effective cross section $\sigma_{\rm eff}$ in the three $\ptsj$ bins.}
\label{tab:sigma_eff}
\begin{tabular}{ccc} \hline\hline
$\ptsj$ GeV ~&~ $\la \ptsj \ra$ (GeV) ~&~ $\sigma_{\rm eff}$ (mb) \\\hline
$15-20$ ~&~ $17.6$ ~&~ $18.2 \pm 3.8$ \\
$20-25$ ~&~ $22.3$ ~&~ $16.3 \pm 3.7$ \\
$25-30$ ~&~ $27.3$ ~&~ $13.9 \pm 4.5$ \\\hline\hline
\end{tabular}
\end{table}

\begin{table*}[t]
\begin{center}
\small
\caption{Systematic ($\delta_{\rm syst}$), statistic ($\delta_{\rm stat}$) and total $\delta_{\rm total}$ uncertainties 
(in \%) for $\sigma_{\rm eff}$ in the three $\ptsj$ bins.}
\label{tab:uncert}
\vskip 1mm
\begin{tabular}{ccccccccc} \hline\hline
 $\ptsj$  & \multicolumn{5}{c}{Systematic uncertainty sources } &  $\delta_{\rm syst}$  & $\delta_{\rm stat}$ &  $\delta_{\rm total}$ \\
 (GeV)    &  ~$f_{\rm DP}$~ & ~$f_{\rm DI}$~ & $\varepsilon_{\rm DP}/\varepsilon_{\rm DI}$ &~ JES ~&~ $R_c \sigma_{\rm hard}$ ~&  (\%) & (\%) & (\%) \\\hline
  15 -- 20 &   7.9 &  17.1 &   5.6 &   5.5 &   2.0 &    20.5 &     3.1 &    20.7 \\\hline
  20 -- 25 &   6.0 &  20.9 &   6.2 &   2.0 &   2.0 &    22.8 &     2.5 &    22.9 \\\hline
  25 -- 30 &  10.9 &  29.4 &   6.5 &   3.0 &   2.0 &    32.2 &     2.7 &    32.3 \\\hline\hline
\end{tabular}
\end{center}
\vskip -3mm
\end{table*}
The measured $\sigma_{\rm eff}$ values in the different $\ptsj$ bins 
agree with each other within their uncertainties, however a slow decrease with $\ptsj$ can not be excluded.
The $\sigma_{\rm eff}$ value averaged over the three $\ptsj$ bins is
\begin{equation}
 \sigma_{\rm eff}^{\rm ave} = 16.4 \pm 0.3(\rm stat) \pm 2.3(\rm syst)  {~~\rm mb}.
  \label{eq:sigeff_av}
\end{equation}

\subsection{Models of parton spatial density}

In this section we study the limits that can be obtained  on the parameters %of a few models 
of three phenomenological models of parton spatial density
using the measured effective cross section (\ref{eq:sigeff_av}).
In the discussion below we follow a simple classical approach.
For a given parton spatial density inside the proton or antiproton $\rho(r)$,
one can define a (time-integrated) overlap $\cal O({\beta})$ 
between the parton distributions of the colliding nucleons 
as a function of the impact parameter $\beta$~\cite{Sjost}.
The larger the overlap (i.e. smaller $\beta$), the more probable it is to have at least one parton interaction
in the colliding nucleons.
The single hard scattering cross sections (for example, \gpj or dijet production) 
should be proportional to $\cal O({\beta})$
and  the cross section for the double parton scattering is proportional to the squared overlap,
both integrated over all impact parameters $\beta$~\cite{Brown,Treleani_2007}:
\begin{equation}
 \sigma_{\rm eff} = \frac{[\int_{0}^{\infty} {\cal O}(\beta) ~2\pi \beta ~{\rm d} \beta]^2 }
                         {\int_{0}^{\infty} {\cal O}(\beta)^2 ~2\pi \beta ~{\rm d} \beta}.
  \label{eq:sigeff_int}
\end{equation}
First, we consider the ``solid sphere'' model
with a constant density inside the proton radius $r_p$.
In this model, the total hard scattering cross section can be written as $\sigma_{\rm hard} = 4\pi r_p^2$
and $\sigma_{\rm eff} = \sigma_{\rm hard} / f$. Here $f$ is the geometrical
{\rm enhancement} factor of the DP cross section. It is obtained by solving Eq.~(\ref{eq:sigeff_int})
for two overlapping spheres with a boundary conditions that the parton density
$\rho(r)= {\rm constant}$ for $r\leq r_{p}$ and $\rho(r)=0$ for $r>r_{p}$ and found to be $f = 2.19$.
The role of the enhancement factor can be seen better if we rewrite Eq.~(\ref{eq:sigma_DPS}) as
$\sigma_{\rm DP} = f {\sigma_{\rm A} \sigma_{\rm B}} / {\sigma_{\rm hard}}$.
The harder the single-parton interaction is the more it is biased 
towards the central hadron-hadron collision with a small impact parameter, where we have a larger
overlap of parton densities and, consequently, higher probability 
for a second parton interaction~\cite{TH3}.
Using the measured  $\sigma_{\rm eff}$, for the solid sphere model we extract 
the proton radius $r_p = 0.53 \pm 0.06$ fm and proton rms-radius $R_{\rm rms} = 0.41\pm 0.05$ fm. 
The latter is  obtained from averaging $r^2$ 
as $R_{\rm rms}^2 \equiv \int_0^{\infty} r^2 4\pi r^2 \rho(r) dr =  4\pi \int_0^{\infty} \rho(r) r^4 dr$~\cite{Hoft}. 
The results are summarized in the line ``Solid Sphere'' of Table~\ref{tab:part_dens}.
\begin{table*}[htbp]                                                                          
\caption{Parameters of parton spatial density models calculated from measured $\sigma_{\rm eff}$.}
\label{tab:part_dens}
\begin{tabular}{cccccccc} \hline \hline
Model for density  & $\rho(r)$ & $\sigma_{\rm eff}$ & $R_{\rm rms}$ & ~~Parameter (fm)~~ & ~~$R_{\rm rms}$ (fm)~~ \\\hline
 Solid Sphere & Constant, $r<r_{p}$      & $4\pi r_p^{2}/2.2$ & $\sqrt{3/5}r_p$ & $0.53\pm0.06$ & $0.41\pm 0.05$  \\
 Gaussian     & $e^{-r^{2}/2a^{2}}$ & $8\pi a^{2}$  & $\sqrt{3}a$ & $0.26\pm0.03$ & $0.44\pm 0.05$  \\
 Exponential  & $e^{-r/b}$         & $28\pi b^{2}$   & $\sqrt{12}b$ & $0.14\pm0.02$ & $0.47\pm 0.06$  \\\hline
\hline
\end{tabular}
\vskip 2mm
\end{table*}
The Gaussian model with $\rho(r) \propto e^{-r^{2}/2a^{2}}$ and
exponential model with $\rho(r) \propto e^{-r/b}$ have been also tested.
The relationships between the scale parameter ($r_p$, $a$ or $b$) and
rms-radius for all the models are given in Table~\ref{tab:part_dens}. 
The relationships between the effective cross section $\sigma_{\rm eff}$ and parameters of the Gaussian
and exponential models are taken from~\cite{Trel_09}, neglecting the terms that represent
correlations in the transverse space.
The scale parameters and rms-radii for both models are also given in Table~\ref{tab:part_dens}.
In spite of differences in the models, the proton rms-radii are in good agreement
with each other, with average values varied as $0.41-0.47$ and with about $12\%$ uncertainty.
On the other hand, having obtained rms-radius from other sources 
(for example, \cite{Belitsky}) 
and using the measured $\sigma_{\rm eff}$, the size of the transverse correlations~\cite{Trel_09} 
can be estimated.

\section{Summary}

We have analyzed a sample of $\gamma + 3$ jets events collected by the D0 experiment 
with an integrated luminosity of about 1~fb$^{-1}$ 
and determined the fraction of events with hard double parton scattering
occurring in a single $p \bar{p}$ collision at $\sqrt{s}=1.96$ TeV.
These fractions are measured in three intervals
of the second (ordered in $p_T$) jet transverse momentum $\ptsj$ 
and vary from $0.466\pm0.041$ at $15 \leq \ptsj \leq 20$~GeV to $0.235\pm0.027$ at $25 \leq \ptsj \leq 30$~GeV.

In the same three $\ptsj$ intervals, we calculate an effective cross section $\sigma_{\rm eff}$, 
a process-independent scale parameter which provides information about the parton spatial 
density inside the proton and define the rate of double parton events. 
The measured $\sigma_{\rm eff}$ values agree for the three $\ptsj$ intervals
with an average $\sigma_{\rm eff}^{\rm ave} = 16.4 \pm 0.3(\rm stat) \pm 2.3(\rm syst)$ mb.
We note that this average value is
in the range of those found in previous measurements~\cite{UA2,CDF93,CDF97}
performed at different  parton interaction energy scales,
and may indicate stable behavior of $\sigma_{\rm eff}$      
with respect to the considered energy scales. 

Using the measured  $\sigma_{\rm eff}$ we have calculated scale parameters and rms-radii
of the proton for three models of the parton matter distribution.

~\\[5mm]
\centerline{\bf Acknowledgements}
~\\[1mm]
% acknowledgement_paragraph_r2.tex                         12/15/09
%
We would like to thank T.~Sj\"ostrand and P.~Skands for very useful discussions.
We also thank the staffs at Fermilab and collaborating institutions, 
and acknowledge support from the 
DOE and NSF (USA);
CEA and CNRS/IN2P3 (France);
FASI, Rosatom and RFBR (Russia);
CNPq, FAPERJ, FAPESP and FUNDUNESP (Brazil);
DAE and DST (India);
Colciencias (Colombia);
CONACyT (Mexico);
KRF and KOSEF (Korea);
CONICET and UBACyT (Argentina);
FOM (The Netherlands);
STFC and the Royal Society (United Kingdom);
MSMT and GACR (Czech Republic);
CRC Program, CFI, NSERC and WestGrid Project (Canada);
BMBF and DFG (Germany);
SFI (Ireland);
The Swedish Research Council (Sweden);
and
CAS and CNSF (China).

\section{Appendix A}

In this measurement we assume that the two parton interactions in the DP \gpTHRj events
can be considered to be independent from each other.
Possible correlation may appear both in momentum space, since the two interactions have
to share the same proton momentum, and at the fragmentation stage.

\begin{figure}[htbp]
\hspace*{-0mm} \includegraphics[scale=0.22,bb=1 10 520 510,clip=true]{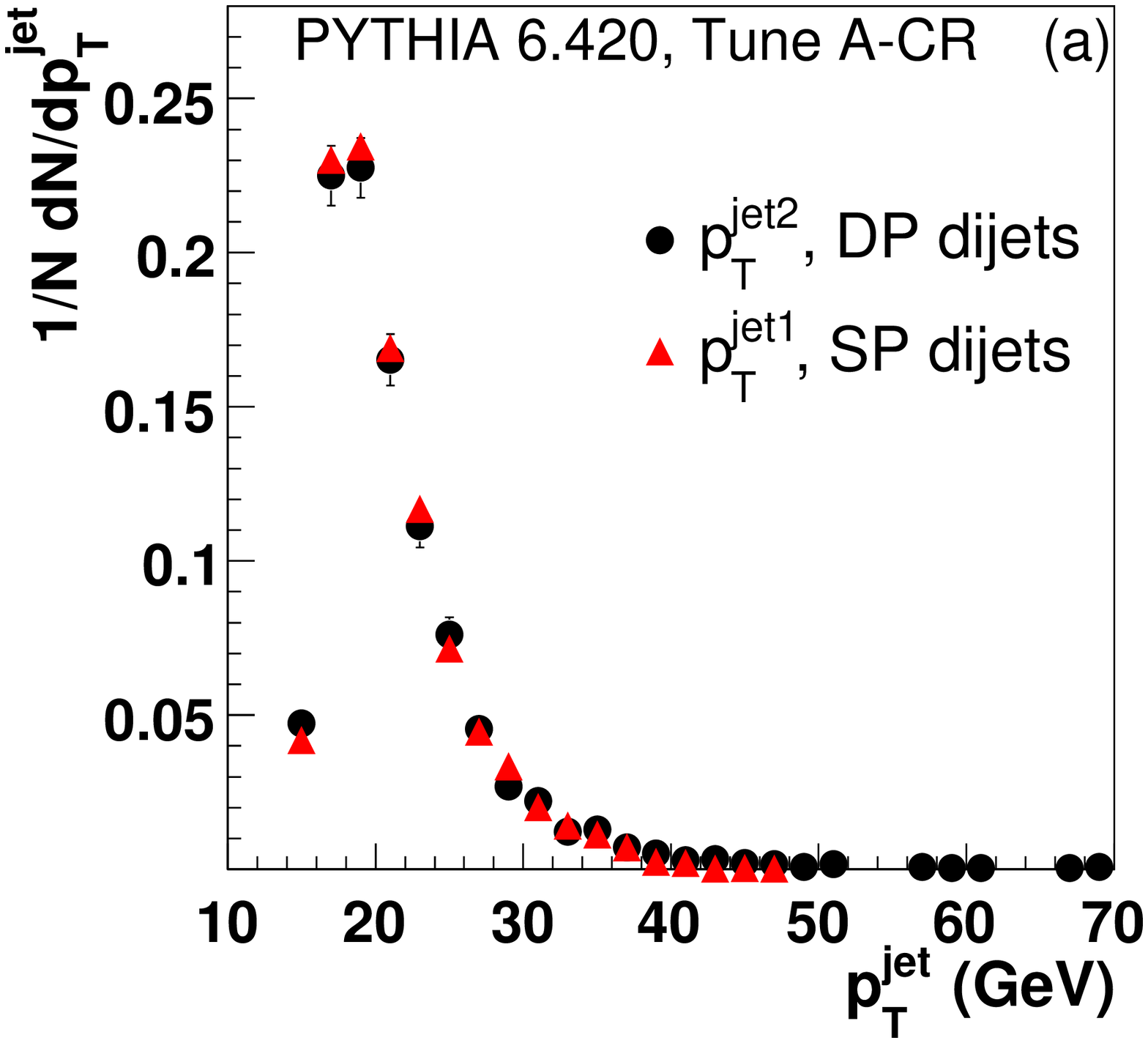}
\hspace*{-0mm} \includegraphics[scale=0.22,bb=1 10 520 510,clip=true]{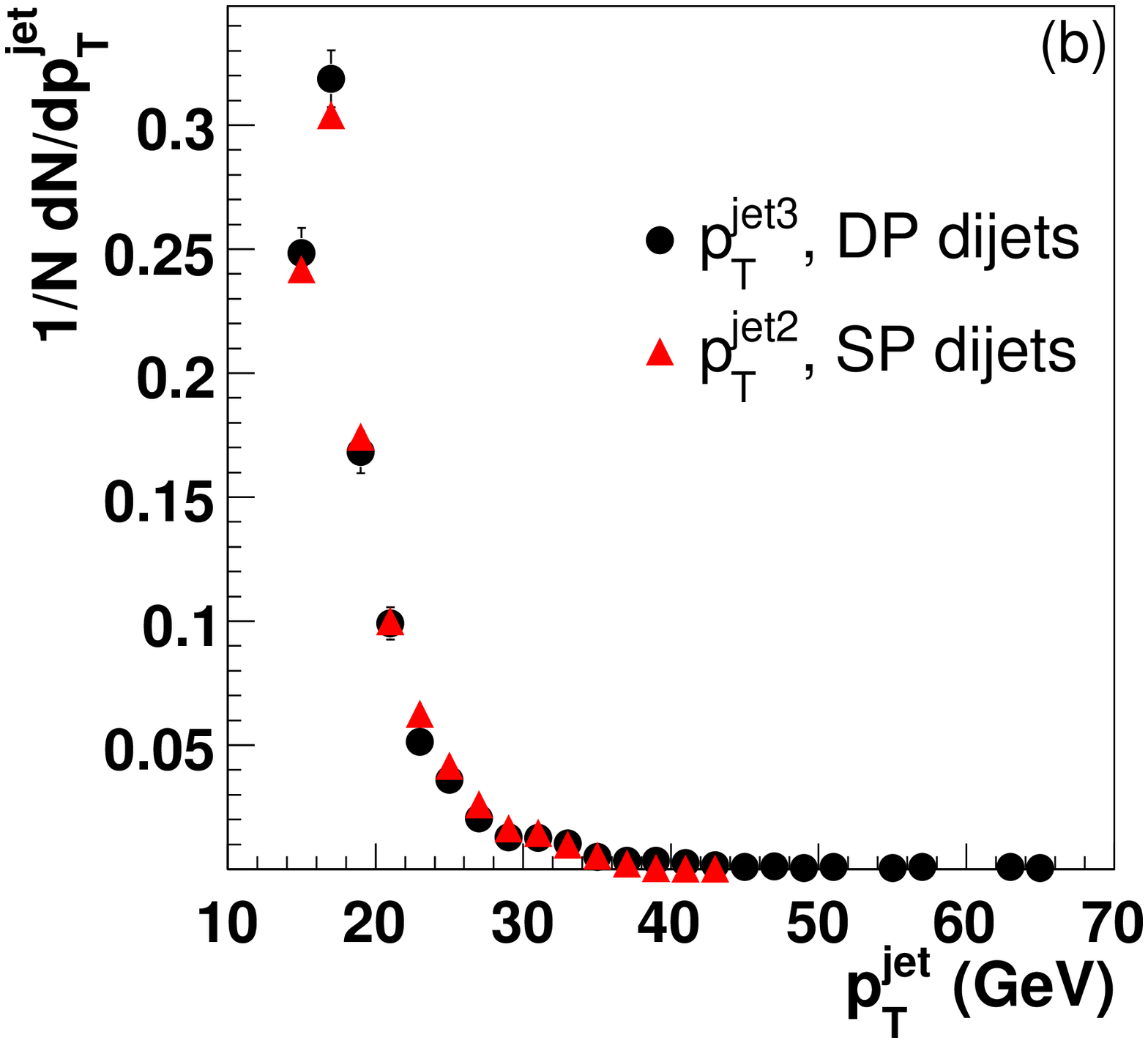}
~\\[2mm]
\hspace*{-0mm} \includegraphics[scale=0.22,bb=1 10 520 510,clip=true]{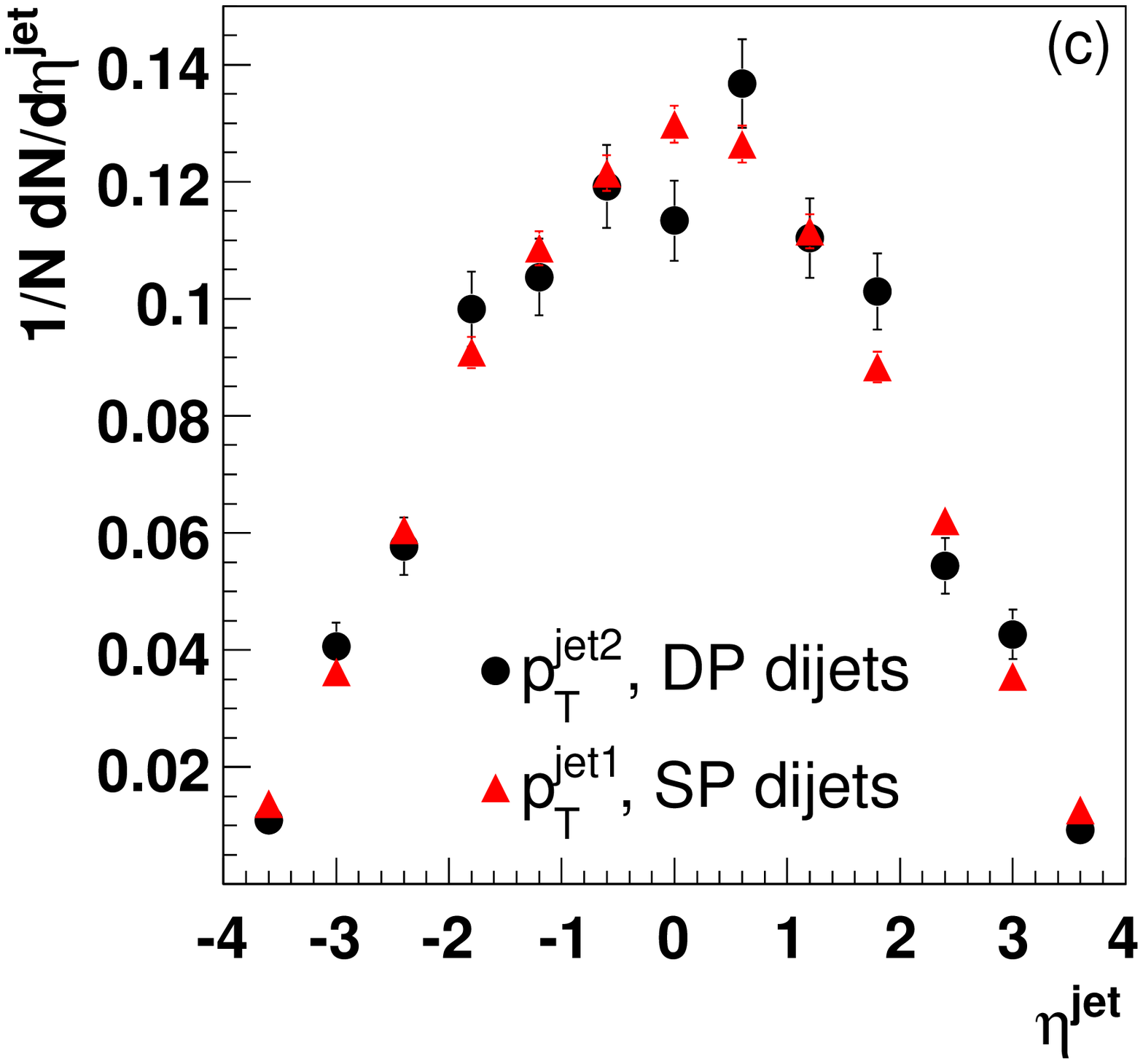}
\hspace*{-0mm} \includegraphics[scale=0.22,bb=1 10 520 510,clip=true]{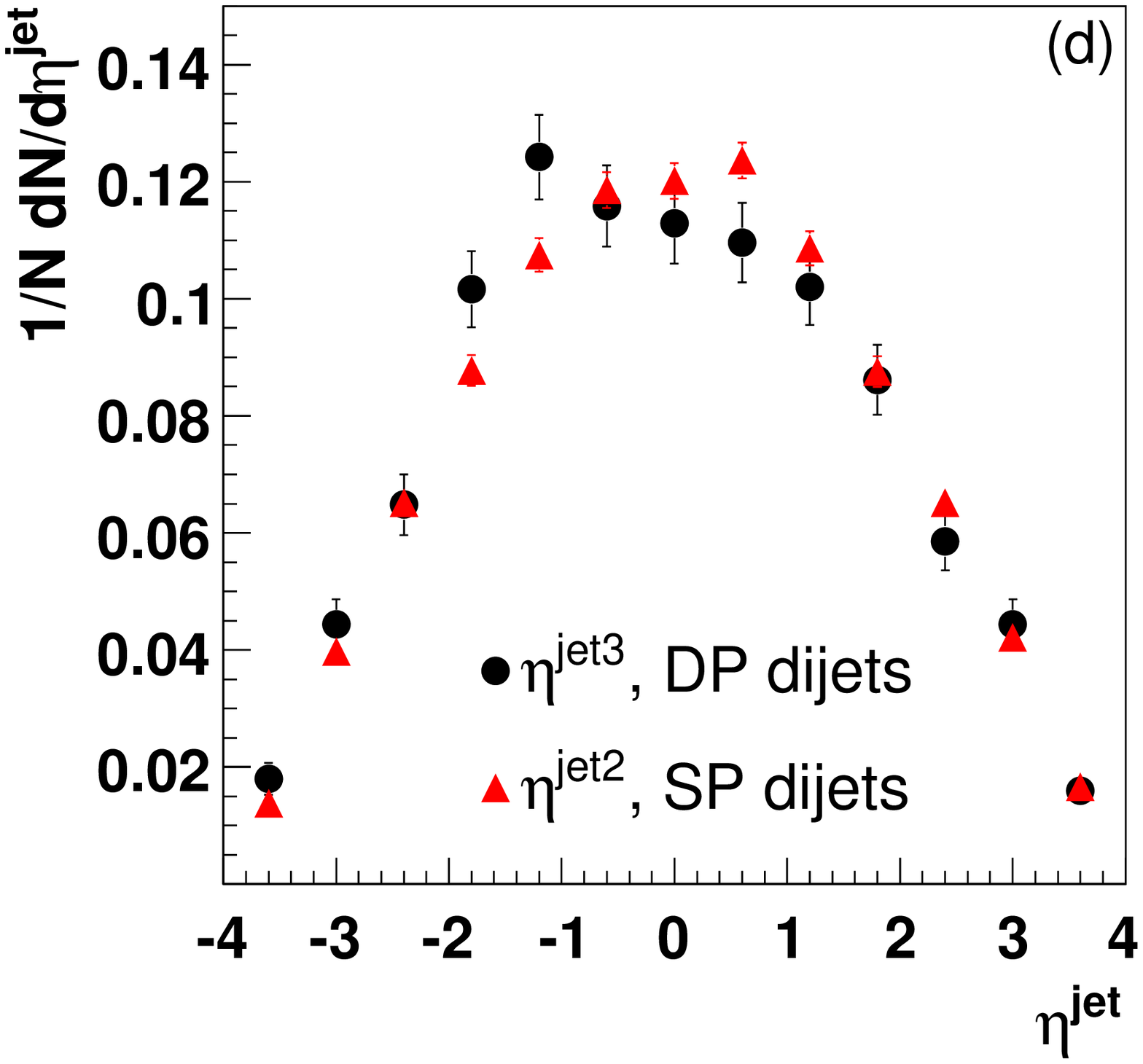}
\caption{Comparison of dijet events properties in SP (triangles) and in \gpTHRj DP events (black circles):
(a) and (c) show comparisons of the $p_T$ and $\eta$ distribution of the second (ordered in $p_T$) jet 
in \gpTHRj DP events with the first jet from the SP dijet events;
(b) and (d) show comparisons of the $p_T$ and $\eta$ distribution of the third jet 
in \gpTHRj DP events with the second jet from the SP dijet events.
Both types of events are generated without ISR and FSR effects but with MPI Tune A-CR.}
\label{fig:pt_tuneACR}
\vskip -3mm
\end{figure}

In the hypothesis of two independent scatterings, the kinematic properties of SP dijet events should
be very similar to those produced in the second parton interaction
in the DP \gpTHRj events.
We compare the $p_T$ and $\eta$ distributions for the two cases
using the {\sc pythia} event generator, which
includes momentum and flavor correlations among the partons participating in MPI. 
It also provides the possibility of choosing different MPI models.
In our comparison we use the {\sc pythia} parameters Tune A-CR,
which is usually considered 
as an example of a model with a strong color reconnection
with an extreme prediction for track multiplicities and/or average hadron $p_T$~\cite{Skands}.
As a model for the DP events, we simulate \gpTHRj events using Tune A-CR
but with ISR and FSR effects turned off and applied all selection criteria as described in Sec.~\ref{Sec:ObjectID}.
This configuration of the event generator guarantees that the two jets produced in addition
to the leading jet (and $\gamma$) in the \gpTHRj final state arise only from additional parton interactions.
The $\Delta S$ distribution for these events is shown in Fig.~\ref{fig:DeltaS_tunes} (by triangles).
The SP dijets events are also generated without ISR and FSR. 
Figure~\ref{fig:pt_tuneACR}(a) compares the $p_T$ spectra of the first (in $p_T$)
jet from the second partonic collision in DP events (second jet in \gpTHRj events) and the first
jet in the SP dijet events, while Fig.~\ref{fig:pt_tuneACR}(b) make analogous comparisons of the next (in $p_T$) jet
in both event types. Figures~\ref{fig:pt_tuneACR}(c) and ~\ref{fig:pt_tuneACR}(d) compare the $\eta$ distributions of these jets. 
We can see good agreement between the kinematics of jets produced in the second parton interaction and those from 
the regular SP dijet events.
Analogous comparisons were performed using Tunes A and S0 with similar good agreement. This indicates
the absence of visible correlations between the two DP scatterings with our selection criteria.

\section{Appendix B}

\begin{figure}[htbp]
~\\[-0mm]
\hspace*{-3mm} \includegraphics[scale=0.21,bb=10 20 500 500,clip=true]{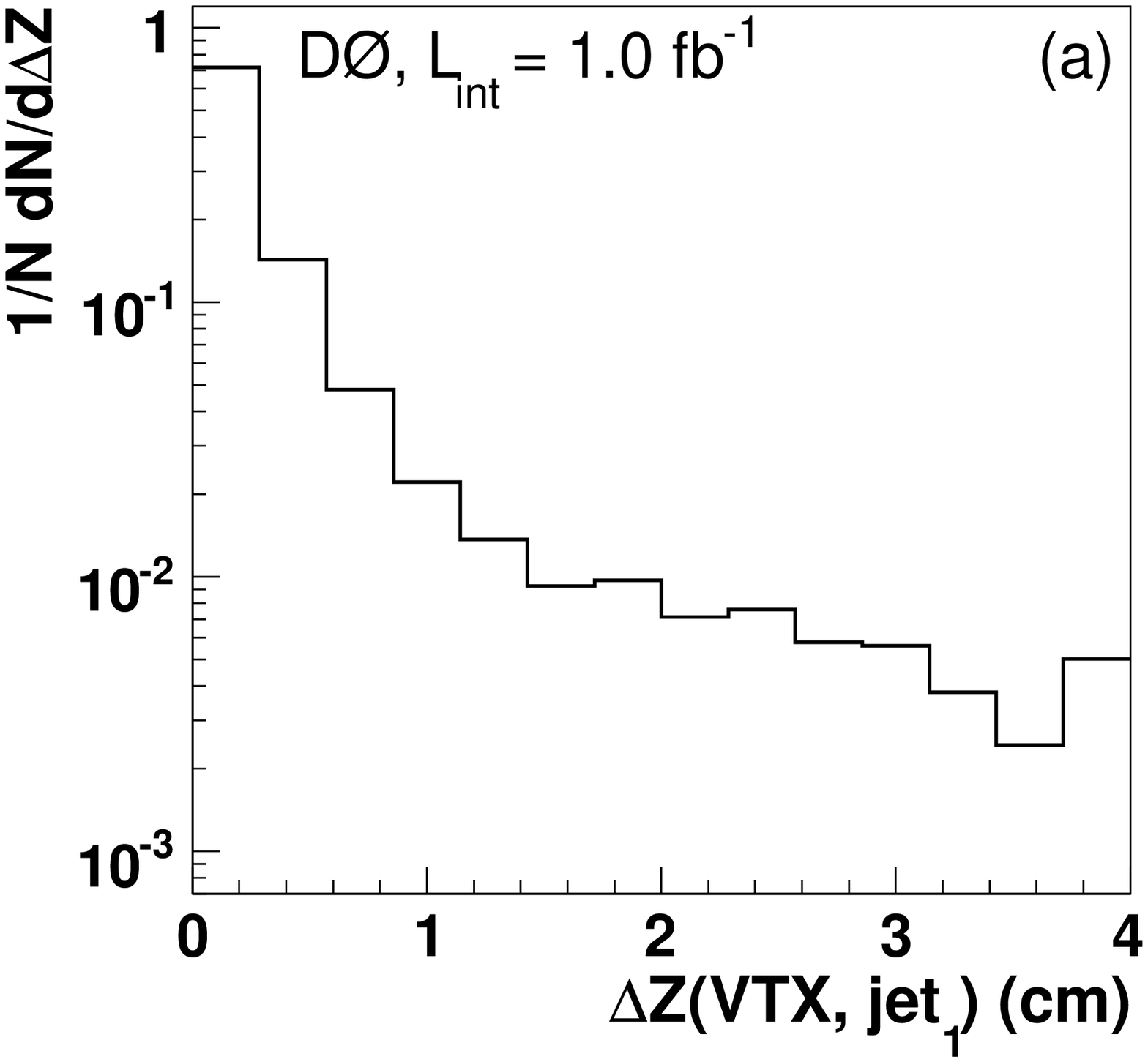}
\hspace*{2mm} \includegraphics[scale=0.21,bb=10 20 500 500,clip=true]{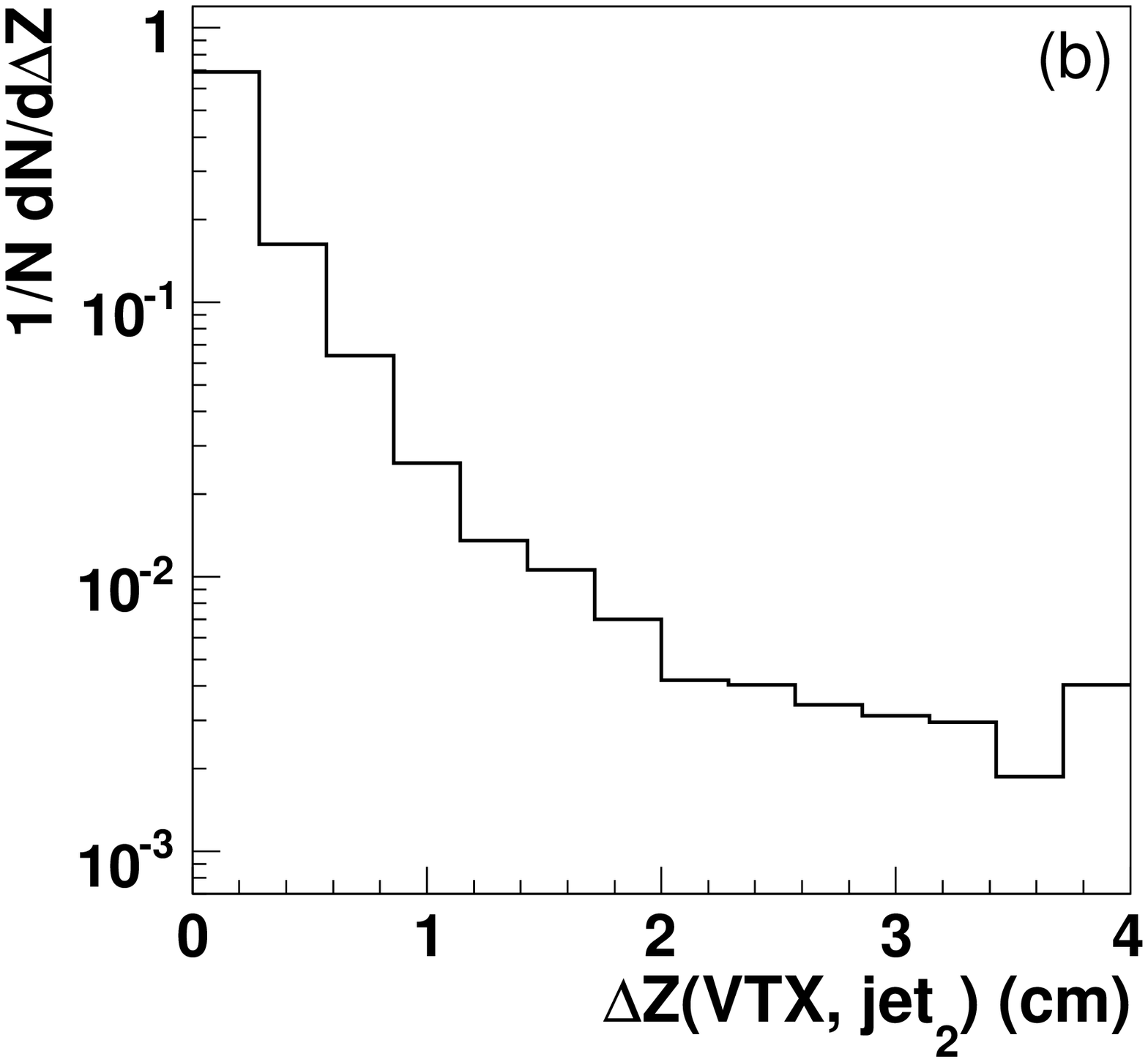}
~\\[2mm]
\hspace*{-2mm} \includegraphics[scale=0.21,bb=10 20 500 500,clip=true]{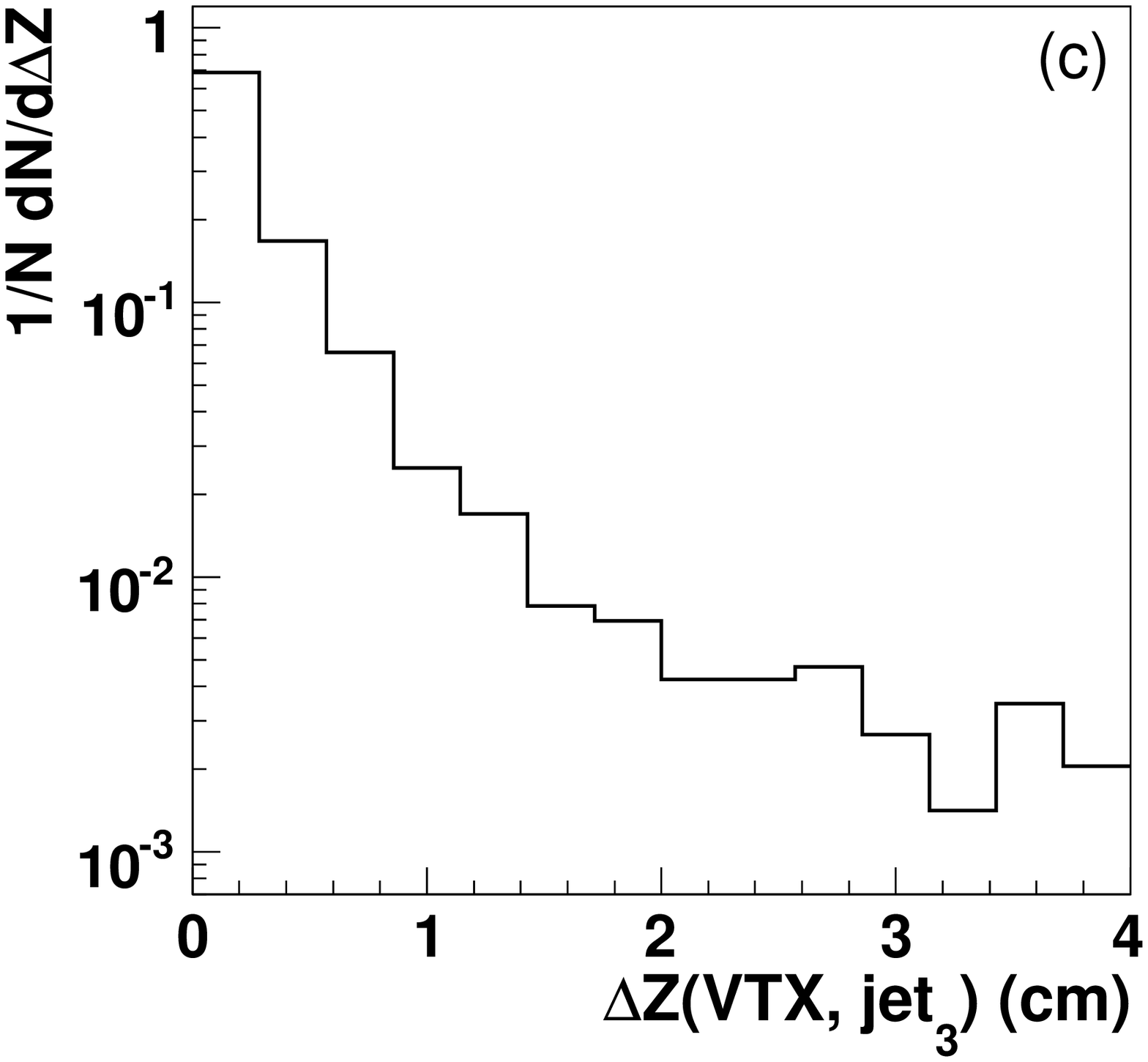}
\vskip -0mm
\caption{Normalized distribution of the number of events as a function of 
the distance along the $z$-axis between
jet$_-z$  (see text) and the $p\bar{p}$ vertex closest in $z$-position
for the (a) first, (b) second and (c) third jets in the 2Vtx data sample.}
\label{fig:jets_dZ_ztrk}
\end{figure}

In building signal and background DI models in Sec.~\ref{Sec:Models},
we take into account information about tracks associated with jets. 
We use two algorithms. In the first, we
consider all tracks inside a jet radius (${\cal R} =0.7$ in our case) 
and calculate the $p_T$-weighted position in $z$ of all the tracks (``jet$_-z$'').
Here the track $z$ position is calculated at the point of closest approach of this track
to the beam ($z$) axis. For each jet in the 2Vtx data sample (Sec.~\ref{Sec:ObjectID})
we can estimate the distance between the  jet$_-z$ and the $p\bar{p}$ vertex closest in $z$, $\Delta z({\rm Vtx,~jet}_i)$. 
These distributions are shown in Fig.~\ref{fig:jets_dZ_ztrk} for each jet in the \gpTHRj 2Vtx sample.
About (95--96)\%~[(97--99\%)] of events have $\Delta z({\rm Vtx,~jet}_i)<1.5~(2.0)$ cm. 

We also use an algorithm that is based on a jet charged particle fraction (CPF) 
and define a discriminant which measures the probability that a given jet originates
from a particular vertex (a jet, having originated from a vertex, may still have tracks coming
from another vertex). 
The CPF discriminant is based on  the fraction of charged transverse energy in each jet $i$ (in the form of
tracks) originating from each identified vertex $j$ in the event:\\[-7mm]
\begin{eqnarray}
{\rm CPF(jet}_i,{\rm Vtx}_j) = \frac{\sum_k p_T({\rm trk}_k^{{\rm jet}_i},{\rm Vtx}_j)}{\sum_n\sum_l p_T({\rm trk}_l^{{\rm jet}_i},{\rm Vtx}_n)}.
\end{eqnarray}

To confirm that a given jet originate from a vertex, we require $\Delta z < 2.0$
and ${\rm CPF} > 0.5$. These requirements being applied to two (or three) jets in the 2Vtx
events allow to build the signal and background DI models described in Section IV.

\bibliography{prl}

\begin{thebibliography}{23}
\expandafter\ifx\csname natexlab\endcsname\relax\def\natexlab#1{#1}\fi
\expandafter\ifx\csname bibnamefont\endcsname\relax
  \def\bibnamefont#1{#1}\fi
\expandafter\ifx\csname bibfnamefont\endcsname\relax
  \def\bibfnamefont#1{#1}\fi
\expandafter\ifx\csname citenamefont\endcsname\relax
  \def\citenamefont#1{#1}\fi
\expandafter\ifx\csname url\endcsname\relax
  \def\url#1{\texttt{#1}}\fi
\expandafter\ifx\csname urlprefix\endcsname\relax\def\urlprefix{URL }\fi
\providecommand{\bibinfo}[2]{#2}
\providecommand{\eprint}[2][]{\url{#2}}


\bibitem{Landsh}
\bibinfo{author}{\bibfnamefont{P.V.}~\bibnamefont{Landshoff}} and
\bibinfo{author}{\bibfnamefont{J.C.}~\bibnamefont{Polkinghorne}},
\bibinfo{journal}{Phys. Rev.} {\bibinfo{volume}D {\bf 18}}, \bibinfo{pages}{3344} (\bibinfo{year}{1978}).

\bibitem{Goebel}
\bibinfo{author}{\bibfnamefont{C.}~\bibnamefont{Goebel}}, %{\sl et al}},
\bibinfo{author}{\bibfnamefont{F.}~\bibnamefont{Halzen}}, and
\bibinfo{author}{\bibfnamefont{D.M.}~\bibnamefont{Scott}},
\bibinfo{journal}{Phys. Rev.} {\bibinfo{volume}D {\bf 22}}, \bibinfo{pages}{2789} (\bibinfo{year}{1980}).


\bibitem{TH1}
%  \bibinfo{author}{\bibfnamefont{E.}~\bibnamefont{Takagi}},
%  \bibinfo{journal}{Phys. Rev. Lett.} \textbf{\bibinfo{volume}{43}},
%  \bibinfo{pages}{1296} (\bibinfo{year}{1979}).
  \bibinfo{author}{\bibfnamefont{N.}~\bibnamefont{Paver}} and
  \bibinfo{author}{\bibfnamefont{D.}~\bibnamefont{Treleani}},
  \bibinfo{journal}{Nuovo Cimento}
  {\bibinfo{volume}A {\bf 70}}, \bibinfo{pages}{215} (\bibinfo{year}{1982});
  \bibinfo{author}{\bibfnamefont{L.}~\bibnamefont{Ametller}}, % {\sl et al}},
  \bibinfo{author}{\bibfnamefont{N.}~\bibnamefont{Paver}}, and
  \bibinfo{author}{\bibfnamefont{D.}~\bibnamefont{Treleani}},
  \bibinfo{journal}{Phys. Lett.} {\bibinfo{volume}B {\bf 169}},
  \bibinfo{pages}{289} (\bibinfo{year}{1986}).
  
  
\bibitem{TH2}
  \bibinfo{author}{\bibfnamefont{B.~} \bibnamefont{Humpert}},
  \bibinfo{journal}{Phys. Lett.} {\bibinfo{volume}B {\bf 131}},
  \bibinfo{pages}{461} (\bibinfo{year}{1983});
  \bibinfo{author}{\bibfnamefont{B.}~\bibnamefont{Humpert}} and
  \bibinfo{author}{\bibfnamefont{R.}~\bibnamefont{Odorico}},
  \bibinfo{journal}{Phys. Lett.} {\bibinfo{volume}B {\bf 154}},
  \bibinfo{pages}{211} (\bibinfo{year}{1985}).

\bibitem{TH3}
\bibinfo{author}{\bibfnamefont{T.}~\bibnamefont{Sj\"ostrand}},
Fermilab Report No. FERMILAB-Pub-85/119-T, 1985;
\bibinfo{author}{\bibfnamefont{T.}~\bibnamefont{Sj\"ostrand}} and
\bibinfo{author}{\bibfnamefont{M. van}~\bibnamefont{Zijl}},
\bibinfo{journal}{Phys. Rev.} {\bibinfo{volume}D {\bf 36}}, \bibinfo{pages}{2019} (\bibinfo{year}{1987}).


\bibitem{AFS}
\bibinfo{author}{\bibfnamefont{T.}~\bibnamefont{Akesson}} 
\bibnamefont{{\sl et~al.}} (\bibinfo{collaboration}{AFS Collaboration}),
\bibinfo{journal}{Z.~Phys.} {\bibinfo{volume}C {\bf 34}}, \bibinfo{pages}{163} (\bibinfo{year}{1987}).

\bibitem{UA2}
\bibinfo{author}{\bibfnamefont{J.}~\bibnamefont{Alitti}}
\bibnamefont{{\sl et~al.}} (\bibinfo{collaboration}{UA2 Collaboration}),
\bibinfo{journal}{Phys. Lett.} {\bibinfo{volume}B {\bf 268}}, \bibinfo{pages}{145} (\bibinfo{year}{1991}).

\bibitem{CDF93}
\bibinfo{author}{\bibfnamefont{F.}~\bibnamefont{Abe}}
\bibnamefont{{\sl et~al.}} (\bibinfo{collaboration}{CDF Collaboration}),
\bibinfo{journal}{Phys. Rev.} {\bibinfo{volume}D {\bf 47}}, \bibinfo{pages}{4857} (\bibinfo{year}{1993}).

\bibitem{CDF97}
\bibinfo{author}{\bibfnamefont{F.}~\bibnamefont{Abe}}
\bibnamefont{{\sl et~al.}} (\bibinfo{collaboration}{CDF Collaboration}),
\bibinfo{journal}{Phys. Rev. Lett.} {\bibinfo{volume} {\bf 79}},
  \bibinfo{pages}{584} (\bibinfo{year}{1997});
\bibinfo{journal}{Phys. Rev.} {\bibinfo{volume}D {\bf 56}}, \bibinfo{pages}{3811} (\bibinfo{year}{1997}).

\bibitem{Sjost}
\bibinfo{author}{\bibfnamefont{T.}~\bibnamefont{Sj\"ostrand}} and
\bibinfo{author}{\bibfnamefont{P.~Z.}~\bibnamefont{Skands}},
\bibinfo{journal}{JHEP} {\bibinfo{volume} {\bf 0403}}, \bibinfo{pages}{053} (\bibinfo{year}{2004}).

\bibitem{Threl}
\bibinfo{author}{\bibfnamefont{G.}~\bibnamefont{Calucci}} and
\bibinfo{author}{\bibfnamefont{D.}~\bibnamefont{Treleani}},
\bibinfo{journal}{Nucl. Phys.} {\bibinfo{volume} {\bf B71}}, \bibinfo{pages}{392} (\bibinfo{year}{1999}).
%\bibinfo{journal}{Phys. Rev.} {\bibinfo{volume}D {\bf 60}}, \bibinfo{pages}{054023} (\bibinfo{year}{1999}).


\bibitem{Snigir}
%\bibinfo{author}{\bibfnamefont{A.M.}~\bibnamefont{Snigirev}},
%\bibinfo{journal}{Phys. Rev.} {\bibinfo{volume} D {\bf 68}}, \bibinfo{pages}{114012} (\bibinfo{year}{2003}).
\bibinfo{author}{\bibfnamefont{V.L.}~\bibnamefont{Korotkikh}} and
\bibinfo{author}{\bibfnamefont{A.M.}~\bibnamefont{Snigirev}},
 \bibinfo{journal}{Phys. Lett.} {\bibinfo{volume}B {\bf 594}},
  \bibinfo{pages}{171} (\bibinfo{year}{2004}).

\bibitem{Stirling09}
\bibinfo{author}{\bibfnamefont{J.R.~}~\bibnamefont{Gaunt}} and
\bibinfo{author}{\bibfnamefont{W.J.~}~\bibnamefont{Stirling}},
\bibfnamefont{Cavendish-HEP-09-20, arXiv:0910.4347 [hep-ph]} (\bibinfo{year}{2009}).
%\bibinfo{journal}{JHEP} {\bibinfo{volume} {\bf 05}}, \bibinfo{pages}{026} (\bibinfo{year}{2006}).

\bibitem{Sjost_JI}
\bibinfo{author}{\bibfnamefont{T.}~\bibnamefont{Sj\"ostrand}} and
\bibinfo{author}{\bibfnamefont{P.~Z.}~\bibnamefont{Skands}},
\bibinfo{journal}{Eur. Phys. J.} {\bibinfo{volume}C {\bf 39}}, \bibinfo{pages}{129} (\bibinfo{year}{2005}).
%Eur. Phys. J. C39 (2005) 129 (LU TP 04-08, hep-ph/0408302)

%\bibitem{Snigir_Dec09}
%\bibinfo{author}{\bibfnamefont{A.M.}~\bibnamefont{Snigirev}},
%\bibinfo{journal}{Submitted to Phys. Rev.} {\bibinfo{volume}D}, arXiv:1001.0104 [hep-ph].


%\bibitem{Trel05}
%\bibinfo{author}{\bibfnamefont{E.}~\bibnamefont{Cattaruzza}},
%\bibinfo{author}{\bibfnamefont{A.}~\bibnamefont{Del Fabbro}},
%\bibinfo{author}{\bibfnamefont{D.}~\bibnamefont{Treleani}},
%\bibinfo{journal}{Phys.Rev.} {\bibinfo{volume} {\bf D72}}, \bibinfo{pages}{034022} (\bibinfo{year}{2005}).
%Phys.Rev.D72:034022,2005. 


\bibitem{WH}
\bibinfo{author}{\bibfnamefont{A.}~\bibnamefont{Del Fabbro}},
\bibinfo{author}{\bibfnamefont{D.}~\bibnamefont{Treleani}},
\bibinfo{journal}{Phys. Rev.} {\bibinfo{volume}D {\bf 61}}, \bibinfo{pages}{077502} (\bibinfo{year}{2000});
\bibinfo{journal}{Phys. Rev.} {\bibinfo{volume}D {\bf 66}}, \bibinfo{pages}{074012} (\bibinfo{year}{2002}).

\bibitem{Huss}
\bibinfo{author}{\bibfnamefont{M.Y}~\bibnamefont{Hussein}},
\bibinfo{journal}{Nucl. Phys. Proc. Suppl.} {\bibinfo{volume} {\bf 174}}, \bibinfo{pages}{55} (\bibinfo{year}{2007});
\bibfnamefont{arXiv:0710.0203 [hep-ph].}


\bibitem{D0_det}
  \bibinfo{author}{\bibfnamefont{V.M.}~\bibnamefont{Abazov}}
  \bibnamefont{{\sl et~al.}} (\bibinfo{collaboration}{D0 Collaboration}),
  \bibinfo{journal}{Nucl. Instrum. Methods Phys. Res.} {\bibinfo{volume}A {\bf 565}}, \bibinfo{pages}{463}
  (\bibinfo{year}{2006}).


\bibitem{Tao}
  \bibinfo{author}{\bibfnamefont{M.}~\bibnamefont{Drees}},
 \bibinfo{author}{\bibfnamefont{T.}~\bibnamefont{Han}},
  \bibinfo{journal}{Phys. Rev. Lett.} {\bibinfo{volume} {\bf 77}},
  \bibinfo{pages}{4142} (\bibinfo{year}{1996}).


\bibitem[{\citenamefont{Abazov {\sl et~al.}}(2008)}]{gamjet_PLB}
  \bibinfo{author}{\bibfnamefont{V.M.}~\bibnamefont{Abazov}}
  \bibnamefont{{\sl et~al.}} (\bibinfo{collaboration}{D0 Collaboration}),
  \bibinfo{journal}{Phys. Lett.} {\bibinfo{volume}B {\bf 666}},
  \bibinfo{pages}{435} (\bibinfo{year}{2008});
  \bibinfo{journal}{Phys. Rev. Lett.} {\bibinfo{volume} {\bf 102}},
  \bibinfo{pages}{192002} (\bibinfo{year}{2009}).



\bibitem{D0_diff}
\bibinfo{author}{\bibfnamefont{B.}~\bibnamefont{Abbott} {\sl et al.}} (\bibinfo{collaboration}{D0 Collaboration}),
\bibinfo{journal}{Phys. Lett.} {\bibinfo{volume}B {\bf 440}}, \bibinfo{pages}{189} (\bibinfo{year}{1998});
%
\bibinfo{author}{\bibfnamefont{B.}~\bibnamefont{Abbott} {\sl et al.}} (\bibinfo{collaboration}{D0 Collaboration}),
\bibinfo{journal}{Phys. Lett.} {\bibinfo{volume}B {\bf 531}}, \bibinfo{pages}{52} (\bibinfo{year}{2002});
%
\bibinfo{author}{\bibfnamefont{V.}~\bibnamefont{Abazov} {\sl et al.}} (\bibinfo{collaboration}{D0 Collaboration}),
\bibinfo{journal}{Phys. Lett.} {\bibinfo{volume}B {\bf 574}}, \bibinfo{pages}{169} (\bibinfo{year}{2003}).

\bibitem{rapidity}
  \bibfnamefont{Rapidity is defined as $y=-(1/2)\ln[(E + p_Z)/(E - p_Z)]$, where
    $E$ is the energy and $p_Z$ is the momentum component along
    the proton beam direction.}


\bibitem{etaphi} \bibfnamefont{Pseudorapidity $\eta$ is defined as
    $\eta=-\ln[\tan(\theta/2)]$, where $\theta$ is the polar angle
    with respect to the proton beam direction.}


\bibitem[{\citenamefont{Zeppenfeld {\sl et~al.}}(1994)}]{Run2Cone}
  \bibinfo{author}{\bibfnamefont{G.C.}~\bibnamefont{Blazey}} \bibnamefont{{\sl et~al.}},
  \bibfnamefont{arXiv:hep-ex/0005012} (\bibinfo{year}{2000}).


\bibitem{MB}
\bibfnamefont{The MB data are collected by requiring only a beam crossing and a coincident signal in the two 
luminosity monitors located at $2.7\!<\!|\eta|\!<\!4.4$ in pseudorapidity \cite{D0_det}.}


\bibitem{PYT}
\bibinfo{author}{\bibfnamefont{T.}~\bibnamefont{Sj\"ostrand} {\sl et~al.}},
\bibinfo{journal}{JHEP} {\bibinfo{volume} {\bf 05}}, \bibinfo{pages}{026} (\bibinfo{year}{2006}).


\bibitem{Minuit}
  \bibinfo{author}{\bibfnamefont{F.}~\bibnamefont{James}},
  \bibfnamefont{CERN Program Library Long Writeup D506} (\bibinfo{year}{1994}).

\bibitem{CDF_excl}
\bibfnamefont{It differs from \cite{CDF97},
where the measured inclusive double parton fractions are corrected 
for the fraction of triple parton interactions,
what makes $\sigma_{\rm eff}$ and then double parton cross section exclusive \cite{Treleani_2007}.}


\bibitem{Treleani_2007}
  \bibinfo{author}{\bibfnamefont{D.}~\bibnamefont{Treleani}},
  %``Double parton scattering, diffraction and effective cross section,''
\bibinfo{journal}{Phys. Rev.} {\bibinfo{volume}D {\bf 76}}, \bibinfo{pages}{076006} (\bibinfo{year}{2007}).
%  [arXiv:0708.2603 [hep-ph]].


\bibitem{HMCMLL}
  \bibinfo{author}{\bibfnamefont{R.}~\bibnamefont{Barlow}} and 
  \bibinfo{author}{\bibfnamefont{C.}~\bibnamefont{Beeston}},
  \bibinfo{journal}{Comput. Phys. Commun.}
  \textbf{\bibinfo{volume}{77}},
  \bibinfo{pages}{219} (\bibinfo{year}{1993}).

\bibitem{GEANT} 
\bibinfo{author}{\bibfnamefont{R.}~\bibnamefont{Brun}} and 
\bibinfo{author}{\bibfnamefont{F.}~\bibnamefont{Carminati}},
\bibfnamefont{CERN Program Library Long Writeup W5013} (\bibinfo{year}{1993}).

\bibitem{Klim}
\bibinfo{author}{\bibfnamefont{S.}~\bibnamefont{Klimenko} {\sl et al}},
\bibinfo{journal}{Fermilab-FN-0741} (\bibinfo{year}{2003}).

\bibitem{CDF_xsec_tot}
\bibinfo{author}{\bibfnamefont{F.}~\bibnamefont{Abe} {\sl et al.}} (\bibinfo{collaboration}{CDF Collaboration}),
\bibinfo{journal}{Phys. Rev.} {\bibinfo{volume}D {\bf 50}}, \bibinfo{pages}{5550} (\bibinfo{year}{1994}).

\bibitem{E811_xsec_tot}
\bibinfo{author}{\bibfnamefont{C.}~\bibnamefont{Avila} {\sl et al.}} (\bibinfo{collaboration}{E811 Collaboration}),
\bibinfo{journal}{Phys. Lett.} {\bibinfo{volume}B {\bf 445}}, \bibinfo{pages}{419} (\bibinfo{year}{1999}).

\bibitem{CDF_xsec-2Dif} 
\bibinfo{author}{\bibfnamefont{T.}~\bibnamefont{Affolder} {\sl et al.}} (\bibinfo{collaboration}{CDF Collaboration}),
\bibinfo{journal}{Phys. Rev. Lett.} {\bibinfo{volume}D {\bf 87}}, \bibinfo{pages}{141802} (\bibinfo{year}{2001}).

\bibitem{SCHUL}
\bibinfo{author}{\bibfnamefont{G.A.}~\bibnamefont{Schuler}} and 
\bibinfo{author}{\bibfnamefont{T.}~\bibnamefont{Sj\"ostrand}},
\bibinfo{journal}{Phys. Rev.} {\bibinfo{volume}D {\bf 49}}, \bibinfo{pages}{2257} (\bibinfo{year}{1994}).


\bibitem{Brown}
D.~Brown, Ph.D. thesis, Harvard University, 1989.

\bibitem{Hoft}
\bibinfo{author}{\bibfnamefont{R.}~\bibnamefont{Hoftstadter}},
\bibinfo{journal}{Rev. Mod. Phys.} {\bibinfo{volume} {\bf 28}}, \bibinfo{pages}{214} (\bibinfo{year}{1956}).

\bibitem{Trel_09}
\bibinfo{author}{\bibfnamefont{G.}~\bibnamefont{Calucci}} and
\bibinfo{author}{\bibfnamefont{D.}~\bibnamefont{Treleani}},
\bibinfo{journal}{Phys. Rev.} {\bibinfo{volume} {\bf D79}}, \bibinfo{pages}{074013} (\bibinfo{year}{2009}).


\bibitem{Belitsky}
\bibinfo{author}{\bibfnamefont{A.V.~}~\bibnamefont{Belitsky}} and
\bibinfo{author}{\bibfnamefont{A.V.~}~\bibnamefont{Radyushkin}},
  %``Unraveling hadron structure with generalized parton distributions,''
\bibinfo{journal}{Phys. Rev.} {\bibinfo{volume} {\bf 418}}, \bibinfo{pages}{1} (\bibinfo{year}{2005}).
%  [arXiv:hep-ph/0504030].

\bibitem{Skands}
D.~Wicke and P.Z.~Skands, arXiv:0807.3248 [hep-ph] (Fig.3).
%P.Z.~Skands, http://home.fnal.gov/$\sim$mskands/leshouches-plots/


\end{thebibliography}


\begin{thebibliography}{99}
% list_of_visitor_addresses_r2.tex                         12/15/09
%  available symbols are:
%  $\ast, \dag, \ddag, \S, \P, $\|$, $\ast\ast$, \dag\dag, \ddag\ddag ,\#
%
\bibitem[a]{alton}
Visitor from Augustana College, Sioux Falls, SD, USA.
\bibitem[b]{burdin}
Visitor from The University of Liverpool, Liverpool, UK.
\bibitem[c]{haas}
Visitor from SLAC, Menlo Park, CA, USA.
\bibitem[d]{juste}
Visitor from ICREA/IFAE, Barcelona, Spain.
\bibitem[e]{luna-garcia}
Visitor from Centro de Investigacion en Computacion - IPN,
  Mexico City, Mexico.
\bibitem[f]{podesta-lerma}
Visitor from ECFM, Universidad Autonoma de Sinaloa, Culiac\'an, Mexico.
\bibitem[g]{weber}
Visitor from Universit{\"a}t Bern, Bern, Switzerland.
%\bibitem[?]{hooper}
%Visitor from Bradley University, Peoria, IL, USA.
%\bibitem[?]{kozminski}
%Visitor from Lewis University, Romeoville, IL, USA.
%\bibitem[\ddag]{deceased}
%Deceased.

%
\vskip 0.25cm

\end{thebibliography}
\bibliographystyle{apsrev}

\end{document}